\documentclass{aa}  

\usepackage{tabularx}
\usepackage{graphicx}
\usepackage{pdflscape}
\usepackage{ragged2e}
\usepackage{txfonts}
\usepackage{orcidlink}
\usepackage{placeins}
\usepackage{natbib,twoopt}

\begin{document}

\title{Revised optical classification of a sample of gamma-ray emitting AGN with GTC and VLT.}

\author{B. Dalla Barba\inst{1,2}\thanks{\email{benedetta.dallabarba@inaf.it}}\orcidlink{0009-0007-1729-2352}
   \and L. Foschini\inst{2}\orcidlink{0000-0001-8678-0324}
   \and M. Berton\inst{3}\orcidlink{0000-0002-1058-9109}
   \and E. Järvelä\inst{4}\orcidlink{0000-0001-9194-7168}
   \and P. Marziani\inst{5}\orcidlink{0000-0002-6058-4912}
   \and S. Ciroi\inst{6}\orcidlink{0000-0001-9539-3940}
   \and E. Dalla Bontà\inst{6,5,7}\orcidlink{0000-0001-9931-8681}
   \and L. Crepaldi\inst{6,8}\orcidlink{0000-0001-7913-0577}
   \and A. Vietri\inst{5,6}\orcidlink{0000-0003-4032-0853}
   \and S. Antón\inst{9}\orcidlink{0000-0002-0658-644X}
   \and M.~J.~M. Marchã\inst{10}\orcidlink{0000-0002-5251-5538}
   \and P. Condò\inst{11}\orcidlink{0009-0005-9183-8662}
   \and C. Mazzucchelli\inst{12}\orcidlink{0000-0002-5941-5214}
}

   	\institute{
			$^1$ Università degli studi dell’Insubria, Via Valleggio 11, Como 22100, Italy; \\
			$^2$ Osservatorio Astronomico di Brera, Istituto Nazionale di Astrofisica (INAF), Via E. Bianchi 46, Merate (LC) 23807, Italy; \\
            		$^3$ European Southern Observatory (ESO), Alonso de Córdova 3107, Vitacura Santiago, Chile; \\
            		$^4$ Department of Physics and Astronomy, Texas Tech University, Box 41051, Lubbock, 79409-1051, TX, USA; \\
        	    		$^5$ Osservatorio Astronomico di Padova, Istituto Nazionale di Astrofisica (INAF), Vicolo dell'Osservatorio 5, 35122 Padova, Italy; \\
			$^6$ Dipartimento di Fisica e Astronomia "G. Galilei", Università degli studi di Padova, Vicolo dell'Osservatorio 3, Padova 35122, Italy; \\
			$^7$ Jeremiah Horrocks Institute, University of Central Lancashire, Preston, PR1 2HE, UK; \\
            		$^8$ Osservatorio Astronomico di Cagliari, Istituto Nazionale di Astrofisica (INAF), Via della Scienza 5, 09047 Selargius, Italy.\\
			$^9$ CFisUC, Departamento de Física, Universidade de Coimbra, 3004-516 Coimbra, Portugal;\\
            		$^{10}$ Dept. of Physics \& Astronomy and Dept. of Computer Science, University College London, Gower Street, London, WC1E 6BT, UK; \\
			$^{11}$ Dipartimento di Fisica, Università di Roma Tor Vergata, Via della Ricerca Scientifica 1, 00133 Roma, Italy; \\
			$^{12}$ Instituto de Estudios Astrofísicos, Facultad de Ingeniería y Ciencias, Universidad Diego Portales, Avenida Ejercito Libertador 441, Santiago, Chile. \\
            }

   \date{Received ...; accepted ...}

  \abstract{We report the results of the optical follow-up of a sample of $\gamma-$ray-emitting active galactic nuclei (AGN). New high-quality optical spectra were obtained using Gran Telescopio Canarias with Optical System for Imaging and low-Intermediate-Resolution Integrated Spectroscopy and the European Southern Observatory Very Large Telescope Unit Telescope 1 with the FOcal Reducer and low dispersion Spectrograph 2. From the analysis of these spectra, we confirmed the previous classification as narrow-line Seyfert 1 (NLS1) for four objects and discovered two new NLS1s, bringing the total number of optically confirmed $\gamma-$NLS1s to 26. We also identified two ambiguous cases between NLS1 and intermediate Seyfert (IS), three IS, one broad-line Seyfert 1, and one Seyfert 2. Based on the new spectra, we calculated black hole masses ranging from $10^{6.25}$ to $10^{9.32}$ M$_{\odot}$, and Eddington ratios spanning 0.05 to 2.07. This reclassification reinforces the scenario in which AGN with relatively small black hole masses are capable of launching powerful relativistic jets, contributing to our broader understanding of $\gamma-$ray-emitting AGN.}

   \keywords{active galaxies -- Seyfert -- jets}
   \authorrunning{B. Dalla Barba et al.}
   \titlerunning{{Revised optical classification of a sample of gamma-ray emitting AGN with GTC and VLT.}}

   \maketitle
   
\section{Introduction}\label{sec_intro}
The active galactic nuclei (AGN) family is proving to be increasingly complex and diverse. A classical distinction exists between jetted and non-jetted AGN \citep{2017A&ARv..25....2P}. Focusing on the jetted class, we can further divide them into three main categories: flat-spectrum radio quasars (FSRQs), BL Lacertae objects (BL Lacs), and jetted-NLS1s, together with their misaligned counterparts. Misaligned AGN are characterized by large viewing angles, while the formers have their jet axis directed toward the Earth (for recent reviews, see, for example, \citealt{2017FrASS...4....6F} and \citealt{2019ARA&A..57..467B}).

The optical spectra of jetted AGN display a wide range of features, from prominent emission lines -- as in the case of FSRQs and $\gamma-$ray NLS1s -- to weak or absent lines, as observed in BL Lacs. The same applies to misaligned AGN. Of particular interest is the distinction between FSRQs and $\gamma-$ray NLS1s. Both can exhibit Balmer and forbidden lines, but NLS1s are classically defined by the presence of weak oxygen lines ([O III]/H$\beta < 3$) and strong iron emission. The main discriminant between NLS1s and FSRQs lies in the full width at half maximum (FWHM) of the H$\beta$ line: in FSRQs, it exceeds 2000 km s$^{-1}$, while in NLS1s it remains below this threshold \citep{1985ApJ...297..166O,1989ApJ...342..224G}.

After the discovery of $\gamma-$NLS1s, many open questions arose or gained new relevance, such as: what is the role of $\gamma-$NLS1s in the blazar sequence, given their small black hole masses? Can $\gamma-$NLS1s help pinpoint the region where $\gamma-$rays are produced in AGN?  The answer to the first of these questions came in 2016, when \citet{2016A&A...591A..98B} suggested that $\gamma-$ray NLS1s represent the low-mass and low-luminosity tail of the blazar distribution. The answer to the second question is more complex, as different studies have presented contrasting results. Some support a central origin of the $\gamma-$rays (in the broad-line region, BLR, or in the dusty torus) -- see, for example, \citet{2008Natur.452..966M,2014Natur.515..376G} -- while others favor a more distant origin (in the narrow-line region, NLR) -- see, for example, \citet{2019Univ....5..199F,2025A&A...698A.320D}. In both cases, $\gamma-$NLS1s serve as a unique laboratory for investigating the AGN jet phenomenon.

The impact of NLS1s on the population of $\gamma-$ray sources is currently limited by the small number of known objects of this type. To address this, \citet{2022Univ....8..587F} selected a sample of 2980 optically identified $\gamma-$ray-emitting sources from the 4th {\it Fermi} Gamma-ray Large Area Telescope ({\it Fermi}-LAT) catalog (4FGL) and searched for available optical spectra in the literature. This led to a reclassification of a significant portion of the sample. However, many of the public observations were carried out in the 1970s-1980s and suffer from low spectral resolution and/or quality, which prevented a reliable classification (ambiguous cases, AMB). For these AMB cases, and the candidate NLS1 found by \citet{2022Univ....8..587F}, we obtained new optical spectra for a total of 22 objects, 6 in the Northern hemisphere and 17 in the Southern hemisphere. Excluding few cases as detailed below, we obtained a final sample of 18 objects (see Table~\ref{tab_obs}). These spectra allowed us to search for additional cases of $\gamma-$NLS1s and to increase the number of known sources in this family. 

At the same time, we identified some cases of ISs, a subclass of Seyfert galaxies characterized by a peculiar spectral shape of the permitted lines \citep{1976MNRAS.176P..61O}. Owing to the intermediate viewing angle, the observer can detect radiation from both the narrow-line and broad-line regions, resulting in a composite line profile with broad wings and a narrow core. These two contributions are usually modeled with separate Gaussian components, referred to as the broad and narrow components, whose combination reproduces the total observed emission. Because of the peculiar spectral shape of H$\beta$, ISs are often misclassified as NLS1s, and vice versa (see, e.g., \citealt{2020A&A...636L..12J,2025A&A...696A..74C}). Disentangling the two classes requires high-resolution spectra, as confirmed by the NLS1/IS cases found in our sample. Moreover, according to the unified model \citep{1980AJ.....85..198K,1993ARA&A..31..473A,1995PASP..107..803U}, the intermediate viewing angle expected for ISs makes their $\gamma-$ray detection particularly puzzling with respect to jet orientation. The inclusion of NLS1s and ISs as $\gamma-$ray emitting AGN will help us investigate the jetted AGN phenomenon in greater detail, including classes of sources that are often overlooked.

The structure of this paper is as follows. In Section~\ref{sec_obs}, we describe the observations and the data analysis. In Section~\ref{sec_crit}, we listed the classification criteria. In Section~\ref{sec_phys}, we presented how we estimated the physical parameters, including the black hole mass (M$_{\rm BH}$) and the Eddington ratio (R$_{\rm Edd}$). In Section~\ref{sec_notes}, we present the notes for the individual analyzed objects. In Section~\ref{sec_disc}, we discuss the new classifications and present the conclusions. Throughout this paper, we adopt a standard $\Lambda$CDM cosmology with H$_0 = 73.3$ km s$^{-1}$ Mpc$^{-1}$, $\Omega_{\rm matter} = 0.3$, and $\Omega_{\rm vacuum} = 0.7$  \citep{2022ApJ...934L...7R}.

\section{Observations and data analysis}\label{sec_obs}
From the original sample of nearly 3000 objects analyzed by \citet{2022Univ....8..587F}, we selected a subsample of 22 sources for which we requested high-quality spectra. The subsample is the result of the selection of those objects classified by \citet{2022Univ....8..587F} as NLS1 or AMB in their paper, taking also into account for the observability of the objects. The objects are divided into two groups: those in the Northern Hemisphere and those in the Southern Hemisphere. 

The new data were collected using the Optical System for Imaging and low-Intermediate-Resolution Integrated Spectroscopy (OSIRIS), mounted on the Gran Telescopio Canarias (GTC) in the Northern Hemisphere (Program GTC33-22B, PI: E. Järvelä), and the FOcal Reducer/low dispersion Spectrograph 2 (FORS2), mounted on Unit Telescope 1 of the Very Large Telescope (UT1/VLT) in the Southern Hemisphere (Programs 110.23UC.003, PI: C. Mazzucchelli; 111.24P0.002, PI: M. Berton). The OSIRIS sample comprises 6 sources. The FORS2 sample initially included 17 objects as part of a filler program designed for observations during non-optimal conditions at Paranal. The UT1 observations cover 12 of the 17 objects; one spectrum corresponds to PKS 2004$-$447 (J2007$-$4432), which was excluded from this analysis as it has already been studied in detail by \citet{2021A&A...654A.125B}, and the other four spectra show only forbidden narrow lines with FWHM dominated by instrumental broadening. As such, they cannot be used to refine the classification or to derive reliable physical parameters. The final sample, consisting of 18 objects, is listed in Table~\ref{tab_obs}.

The GTC observations were carried out with the following setup: a 1" slit and the R1000R grism, yielding a spectral resolution of $R = \lambda/\Delta \lambda = 1122$\footnote{\url{https://www.gtc.iac.es/instruments/osiris+/osiris+.php\#Longslit_Spectroscopy}}. Standard stars were observed with a 2.52" slit. The OSIRIS spectral coverage is 3650-10000\AA. The UT1/VLT setup was: a 1" slit and the 300V grism, resulting in a spectral resolution of $R = 440$\footnote{\url{https://www.eso.org/sci/facilities/paranal/instruments/fors/overview.html}}. The FORS2 spectral coverage is 3300-11000\AA. Further details are provided in Table~\ref{tab_obs}.

The reduction of the GTC data was carried out with a standard procedure using {\tt IRAF (version 2.18)} (\citealt{1986SPIE..627..733T,1993ASPC...52..173T}). For VLT observations, we used the already reduced data that can be found in the European Southern Observatory (ESO) archive\footnote{\url{https://archive.eso.org/scienceportal/home}}. 

Due to the different slit apertures used for the sources and the standard stars in the GTC observations, a flux correction was applied to the final spectra. Next, for each spectrum, we corrected for reddening and identified the main optical emission lines to compute the redshift ($z$). The redshift was calculated as the flux-weighted average of the individual $z$ values derived from the [O II]$\lambda$3727, H$\beta$, and [O III]$\lambda\lambda$4959,5007 lines, when present. In one case, J2354$-$0958, the measured redshift differed significantly from the value reported in the literature; see Section~\ref{secJ2354} for details. The resulting redshifts are listed in Table~\ref{tab_fit}. Continuum subtraction was performed using the dedicated {\tt IRAF} function. Iron multiplets emission lines were removed using the tool developed by \citet{2010MSAIS..15..176K} and \citet{2012ApJS..202...10S}, available through the Serbian Virtual Observatory\footnote{\url{http://servo.aob.rs/FeII_AGN/}}. The iron fitting allows us to calculate the $R_{4570}$ parameter, defined as the ratio between the flux of the blueward iron bump (relative to H$\beta$) and the flux of the H$\beta$ line.

After these preprocessing steps, we fitted the H$\beta$--[O III] complex, where feasible. In some spectra, the region around H$\beta$ and [O III] was too noisy (e.g., J0102$+$4214), or significantly affected by atmospheric absorption (e.g., J0224$+$0700). In such cases, we fit only the [O III]$\lambda\lambda$4959,5007 lines, or alternatively the Mg II $\lambda$2800 line. We did not perform a telluric correction because the resulting spectral profile cannot be completely trusted to calculate the FWHM and to propose a reliable classification of the AGN. 

The H$\beta$ line was modeled using four alternative profiles: three Gaussians, two Gaussians, one Gaussian, or a Lorentzian. The first two profiles are typically attributed to IS, when the broad and the narrow component are clearly separable, but they can also fit well NLS1 in low-/intermediate-resolution spectra \citep{2023Physi...5.1061D}. The single Gaussian with a FWHM comparable to the one of narrow lines is typical of type 2 AGN. The Lorentzian profile, instead, is typical of NLS1 \citep{2002ApJ...566L..71S,2009NewAR..53..198S,2012MNRAS.426.3086G,2016MNRAS.462.1256C,2020CoSka..50..270B}. 

For the [O III]$\lambda\lambda$4959,5007 doublet, a single Gaussian was sufficient in most cases; however, a two-Gaussian model was adopted when required. This composite model includes a narrow core component from the NLR and an additional outflow component. The two [O III] lines were tied in terms of FWHM, central wavelength, and flux, assuming the theoretical intensity ratio of (2.993$\pm$0.014) \citep{2007AIPC..895..313D}. Parameter uncertainties were estimated using a Monte Carlo method creating $N=1000$ synthetic spectra. For each iteration, we added a Gaussian noise to the line profile, proportional to the standard deviation of the signal in the continuum between 5050--5150\AA. 

\begin{table*}
\centering
\small
\caption{List of the sources with the observation details.}
	\begin{tabular}{llllllllll}
            \hline
		 {\bf Name} &  {\bf Alias} &  {\bf RA (J2000)} &  {\bf DEC (J2000)} & {\bf z$_{pub}$} &  {\bf Tel.} &  {\bf Obs. Date} & {\bf Exp. time} & {\bf Airmass}& {\bf Seeing}\\
		 {\bf } &  {\bf } &  {\bf [h:m:s]} &  {\bf [$^{\circ}$:':'']} &  {\bf } &  {\bf } &  {\bf [yyyy-mm-dd]} & {\bf [s]} & {\bf} & {\bf ["]}\\
            \hline
            \hline
		J0102$+$4214	& 	GB6 J0102$+$4214    	& 01:02:27.1	&   $+$42:14:18.9	&	0.874	&	GTC	&	2022-09-04	& 600 	& 1.35	&	-\\ 
		J0224$+$0700 	&    	PKS 0221$+$067     		& 02:24:28.4  	&   $+$06:59:23.3	&      0.511	&	GTC	&      2022-09-04	& 600	& 1.08	&	-\\         
		J0422$-$0644	&	PMN J0422$-$0643    	& 04:22:10.8   	&   $-$06:43:45.3	&     	0.242	&	UT1	&     	2023-01-25	& 1800 	& 1.88	&	0.71\\              
		J0442$-$0017	&	PKS 0440$-$00      		& 04:42:38.7  	&   $-$00:17:43.4	&      	0.844	&	UT1	&      2023-07-19 	& 1830 	& 1.94	&	0.53\\    
		J0515$-$4556 	&	PKS 0514$-$459    		& 05:15:45.3 	&   $-$45:56:43.2	&      	0.194	&	UT1	&      2022-11-25  	& 1800 	& 1.13	&	0.85\\        
		J0521$-$1734 	&	TXS 0519$-$176      		& 05:21:23.6    	&   $-$17:37:30.1	&      	0.347	&	UT1	&     	2022-12-03 	& 1800 	& 1.07	&	1.65\\     
		J0932$+$5306 	&    	S4 0929$+$53       		& 09:32:41.2     &   $+$53:06:33.8	&      	0.597	&	GTC	&     	2022-10-30 	& 300	& 1.30	&	-\\         
		J1048$-$1912 	&	PKS 1045$-$18       		& 10:48:06.6     &   $-$19:09:35.7	&    	0.595	&	UT1	&     	2022-12-17 	& 1800 	& 1.10	&	2.08\\      
		J1102$+$5251 	&    	GB6 J1102$+$5249   	& 11:02:49.8   	&   $+$52:50:12.7	&      0.690	&	GTC	&     	2022-11-16  	& 600 	& 1.26	&	-\\              
		J1154$+$4037 	&   	B3 1151$+$408     		& 11:53:54.7    	&   $+$40:36:52.6	&     	0.923	&	GTC	&     	2022-11-24 	& 600 	& 1.35	&	-\\      
		J1202$-$0528  	&    	PKS 1200$-$051    		& 12:02:34.2     &   $-$05:28:02.5	&      0.381	&	UT1	&     	2023-01-20  	& 1800  	& 1.17	&	0.55\\       
		J1246$-$2548 	&    	PKS 1244$-$255   		& 12:46:46.8    	&   $-$25:47:49.3	&     	0.638	&	UT1	&     	2023-07-10   	& 1830	& 1.64 	&	1.77\\   
		J1310$+$5514 	&    	TXS 1308$+$554     		& 13:11:03.2     &   $+$55:13:54.3	&     	0.926	&	GTC	&    	2022-11-28  	& 900 	& 1.57	&	-\\        
		J1331$-$1325 	&	PMN J1331$-$1326    	& 13:31:20.4     &   $-$13:26:05.6	&     	0.250	&	UT1	&    	2023-07-15    	& 1830 	& 1.26	&	0.66\\           
		J1818$+$0903	&   	NVSS J181840$+$090346 &18:18:40.1    	&   $+$09:03:46.2	&      0.354	&	UT1 	&     	2023-06-22  	& 1830	& 1.20	&	0.58\\         
		J1902$-$6748 	&    	PMN J1903$-$6749   	&  19:03:01.2    &   $-$67:49:35.9	&      0.255	&	UT1	&      2023-06-19   	& 1830 	& 1.37	&	1.24\\     
		J2325$-$3559  	&    	CTS 0490      			&  23:25:28.6 	&   $-$35:57:54.2	&      0.360	&	UT1	&     	2023-07-19 	& 1830 	& 1.04	&	1.72\\        
		J2354$-$0958  	&    	PMN J2354$-$0957   	&  23:54:05.5	&   $-$09:57:48.8	&      0.272	&	UT1	&      2023-06-28   	& 1830 	& 1.47	&	1.08\\    
            \hline
	\end{tabular}
	\label{tab_obs}
	\tablefoot{The seeing parameter is the average of the values recorded at the beginning and at the end of the observation. It is reported only for the UT1/VLT observations, as it is not available in the headers of the GTC data.}
\end{table*}

\section{Classification criteria}\label{sec_crit}
As already introduced in Section~\ref{sec_intro}, the optical classification of jetted AGN is not always straightforward, especially when dealing with low-SNR spectra. This is particularly true for ISs and NLS1s, which are often prone to misclassification. The difficulty mainly arises from the similarity in the observed H$\beta$ line profiles when its FWHM is close to 2000 km s$^{-1}$, the conventional threshold adopted to distinguish NLS1s from broader-lined sources (see, for example, figure 2 in \citealt{2023Physi...5.1061D}).

Indeed, a Lorentzian profile -- commonly used to model the permitted lines in NLS1s -- can resemble a composite of two or more Gaussian components, typically employed for ISs (and vice versa). This degeneracy in fitting can lead to ambiguous classifications, particularly in low-resolution or noisy spectra. From a physical standpoint, the Lorentzian shape is generally associated with turbulent motions in the BLR, while the Gaussian components are interpreted as arising from clouds in Keplerian motion within the BLR (e.g., \citealt{2011Natur.470..366K,2012MNRAS.426.3086G}). This highlights that the choice of model to fit the H$\beta$ line is not merely a technical matter but reflects underlying physical conditions. Therefore, the distinction between ISs and NLS1s is not only observational but also tied to the kinematics and structure of the BLR, and cannot be attributed solely to orientation effects.

Furthermore, the H$\beta$ line profile can exhibit marked asymmetries. Such asymmetries may arise from a variety of physical processes. One possibility is the presence of an outflow, often modeled as a moving bubble of ionized gas, which can produce blue- or red-asymmetric line wings (e.g., \citealt{1992ApJS...80..109B}). Another potential explanation involves double-peaked emission lines, which can originate from a rotating Keplerian disk around the central black hole (e.g., \citealt{1990A&A...229..313D,1989ApJ...339..742C}). Finally, asymmetric and systematically shifted line profiles have been proposed as possible signatures of close binary supermassive black holes, where the orbital motion of the BLRs around the two black holes induces observable Doppler shifts (e.g., \citealt{1996ApJ...464L.107G}). While these more complex scenarios are fascinating, they typically manifest in distinctly double-peaked broad-line profiles, which are not observed in the present sample. Therefore, such interpretations can be reasonably excluded in the context of this work. Nonetheless, the variety of possible H$\beta$ profiles illustrates the necessity for careful spectral modeling and a multi-faceted classification approach when dealing with jetted AGN.

These considerations underscore the intrinsic challenges in providing a conclusive optical classification for AGN, particularly when relying solely on spectral features that can be affected by noise, resolution, and modeling degeneracies. Despite these limitations, in the following we outline the empirical criteria adopted to classify the objects in the present sample. These criteria aim to balance the need for physical consistency with the constraints imposed by the available data:

\begin{itemize}
    \item NLS1s: Sources showing a Lorentzian H$\beta$ profile or a multiple-Gaussian fit with smooth wings (i.e., no distinct separation between the NLR and the BLR), with a total FWHM $<2000$ km s$^{-1}$. These spectra typically exhibit weak [O III] lines, with [O III]/H$\beta<3$, and often show the presence of optical Fe II multiplets.
    
    \item Broad-Line Seyfert 1 (BLS1s): Classical type 1 AGN characterized by a broad H$\beta$ line with a total FWHM $>2000$ km s$^{-1}$. The H$\beta$ profile may be fit by either a single Gaussian or multiple components, but the broadness is the defining feature.
    
    \item ISs: AGN showing multiple-Gaussian H$\beta$ profiles with a clear separation between the NLR and BLR components. Unlike NLS1s, no strict cut is imposed on the [O III]/H$\beta$ flux ratio or on the presence of Fe II lines. This class includes Seyfert 1.2--1.9 types and generally exhibits intermediate obscuration between type 1 and type 2 AGN.
    
    \item Seyfert 2 (SY2): Sources displaying only narrow emission lines, typically with FWHM $\sim1000$ km s$^{-1}$. These are associated with type 2 AGN and are interpreted as systems viewed edge on, such that the BLR is obscured or totally absent (see the case of true Sy2, e.g. \citealt{2003cxo..prop.1521P,2012MNRAS.426.3225B,2013MNRAS.433.1764M} and references therein).
    
    \item Ambiguous (AMBs): A heterogeneous class reported by \citet{2022Univ....8..587F}, including objects for which no reliable classification could be assigned. These cases lack crucial information such as published optical spectra, measurable H$\beta$ FWHM, or reliable positional cross-identification. Furthermore, the viewing angle remains unconstrained.
\end{itemize}

\section{Physical parameters}\label{sec_phys}
\subsection{Black hole mass (M$_{BH}$)}\label{sec_mbh}
To estimate the black hole mass, we adopted equations 38 and 40 from \citet{2020ApJ...903..112D}, which relates $M_{\rm BH}$ to the luminosity and line width of H$\beta$ -- specifically, the second-order moment ($\sigma_{\rm H\beta}$) or the FWHM in the case of Lorentzian profiles. As noted by \citet{2020ApJ...903..112D}, the second-order moment is generally preferred over FWHM as it yields tighter correlations with black hole mass. Furthermore, these relations avoid the intermediate step involving the continuum luminosity at 5100\AA, which is a significant advantage when analyzing jetted sources. In such cases, jet emission can contaminate the continuum, leading to an overestimation of the black hole mass.

The adopted equations are:

\begin{equation}
\begin{split}
{\rm log}\left( \frac{M_{\rm BH}}{M_{\odot}}\right)={\rm log}(f)&+7.530+0.703\times\left[{\rm log}\left(\frac{L_{\rm H\beta_{broad}}}{\rm erg~s^{-1}}\right)-42\right]+\\
						&+2.183\times\left[{\rm log}\left(\frac{\sigma_{\rm H\beta}}{\rm km~s^{-1}}\right)-3.5\right]
\end{split}
\end{equation} 

\begin{equation}
\begin{split}
{\rm log}\left( \frac{M_{\rm BH}}{M_{\odot}}\right)={\rm log}(f)&+7.015+0.784\times\left[{\rm log}\left(\frac{L_{\rm H\beta_{broad}}}{\rm erg~s^{-1}}\right)-42\right]+\\
						&+1.387\times\left[{\rm log}\left(\frac{\rm FWHM(H\beta)}{\rm km~s^{-1}}\right)-3.5\right]
\end{split}
\end{equation} 

\noindent where $\sigma_{\rm H\beta}$ is the second-order moment. The relations show a scatter of 0.309 dex and 0.371 dex, respectively. The geometrical factor adopted is the total value reported by \citet{2006A&A...456...75C}, in order to avoid assumptions about the population nature of our sample. It corresponds to $f = (3.85 \pm 1.15)$. $L_{\rm H\beta_{broad}}$ denotes the luminosity of the broad component of the H$\beta$ line.

For sources in which the H$\beta$ line is not detected -- either due to absorption or to spectral noise -- we estimated the black hole mass using the Mg II$\lambda2800$ or [O III]$\lambda5007$ lines. Relations based on Mg II$\lambda2800$ suffer from the same issue as H$\beta$ when the continuum contribution at 3000\AA\ is included. For this reason, we adopted the relation from \citet{2012MNRAS.427.3081T}:

\begin{equation}
\label{eq:M_bh_o3}
\begin{split}
{\rm log}\left( \frac{M_{\rm BH}}{M_{\odot}}\right) = 6.83+0.5&\times{\rm log}\left(\frac{L_{\rm{Mg~II}}}{10^{42}~{\rm erg~s^{-1}}}\right)\\
					&+2\times{\rm log}\left[\frac{{\rm FWHM(Mg~II)}}{10^3~{\rm km~s^{-1}}}\right]
\end{split}
\end{equation}

\noindent which relies solely on the luminosity and FWHM of the Mg II$\lambda2800$ line, thus avoiding the use of the continuum luminosity. The scattering for this relation is of 0.13 dex.\\

The relations based on [O III], however, are known to provide less reliable results, and we therefore treat the corresponding mass estimates as upper limits. The tendency of [O III] to overestimate black hole masses in active galaxies has already been discussed by \citet{2014ApJ...789...17H}. For these cases, we used the revised $M-\sigma_\star$ relation from \citet{2014ApJ...789...17H}:

\begin{equation}
{\rm log}\left( \frac{M_{\rm BH}}{M_{\odot}}\right) = 8.49^{+0.052}_{-0.107}+(4.38\pm0.29)\times{\rm log}\left(\frac{\sigma_{\star}}{200~{\rm km~s^{-1}}}\right)
\label{eq_Mbh_sigma}
\end{equation}

\noindent where $\sigma_{\star} = {\rm FWHM([O~III]_{\rm core})} / 2.35$, and FWHM([O III]$_{\rm core}$) is the width of the core component only. The limitation of this method is evident in Figure 4 of \citet{2005ApJ...627..721G}, where the data show a significant scatter around the relation. The results for all sources are reported in Table~\ref{tab_fit}.

\subsection{Eddington ratio (R$_{Edd}$)}\label{sec_redd}
The Eddington ratio is calculated as:

\begin{equation}
R_{\rm Edd} = \frac{L_{\rm bol}}{L_{\rm Edd}} = \frac{L_{\rm bol}}{1.3\times10^{38}\,(M_{\rm BH}/M_{\odot})}
\end{equation}

The black hole mass used is the one derived in the previous paragraph. The bolometric luminosity ($L_{\rm bol}$), on the other hand, can be estimated using the classical relation $L_{\rm bol} = 9 \times \lambda L_{\lambda}(5100\text{\AA})$ from \citet{2000ApJ...533..631K}, or with a revised version from \citet{2019MNRAS.488.5185N}:

\begin{equation}
L_{\rm bol} = k_{\rm bol} \times \lambda L_{\lambda}(5100\text{\AA}) \quad ; \quad 
k_{\rm bol} = 40 \cdot \left[\frac{\lambda L_{\lambda}(5100\text{\AA})}{10^{42}~{\rm erg~s^{-1}}}\right]^{-0.2}
\end{equation}

\noindent where $\lambda L_{\lambda}(5100\text{\AA})$ is the continuum luminosity at 5100\AA. The bolometric correction factor $k_{\rm bol}$ should also be adjusted for source inclination. For type~1 AGN (typical inclination $\sim56^{\circ}$), the correction factor is $\sim$1.4, while for face-on accretion disks it is $\sim$2.5 \citep{2019MNRAS.488.5185N}. Due to the lack of reliable inclination information, we adopt an average value of 2, as done by \citet{2025A&A...696A..74C}.

To determine $L_{\rm bol}$, we first need the continuum luminosity at 5100\AA. This parameter can be directly measured from the spectrum in non-jetted sources; however, in jetted AGN, the jet contributes significantly to the continuum, making it difficult to disentangle from the thermal emission of the accretion disk. In such cases, line properties must instead be used. To this end, we applied the relation from \citet{2020ApJ...903..112D}, which connects $\lambda L_{\lambda}(5100\,\text{\AA})$ to $L_{{\rm H}\beta,\,{\rm broad}}$:

\begin{equation}
\begin{split}
\log \left[\lambda L_{\lambda}(5100\text{\AA})\right] = (43.396&\pm0.018)+(1.003\pm0.022)\times \\
		& \times \left[\log\left({\frac{L_{\rm H\beta_{\rm broad}}}{{\rm erg~s^{-1}}}}\right)-41.746\right]
\end{split}
\end{equation}

\noindent This approach relies entirely on the properties of the H$\beta$ line. 

For objects in which H$\beta$ is undetected or affected by absorption, we used either the Mg II $\lambda2800$ or [O III]$\lambda5007$ lines. For the Mg II line, we used an analogous procedure to that applied for H$\beta$, calculating the expected continuum luminosity at 3000\AA, $\lambda L_{\lambda}(3000\AA)$, from the line properties. Combining equations~8 and~7 from \citet{2012MNRAS.427.3081T}, we obtain:

\begin{equation}
\log \left[\frac{\lambda L_{\lambda}(3000\text{\AA})}{10^{44}~{\rm erg~s^{-1}}}\right] = 0.08+0.81\times\log\left({\frac{L_{\rm Mg~II}}{10^{42}~{\rm erg~s^{-1}}}}\right)
\end{equation}

\noindent whose result can be inserted in the corresponding relation for the Eddington ratio of \citet{2019MNRAS.488.5185N}:

\begin{equation}
L_{\rm bol} = k_{\rm bol} \times \lambda L_{\lambda}(3000\text{\AA}) \quad ; \quad 
k_{\rm bol} = 19 \cdot \left[\frac{\lambda L_{\lambda}(3000\text{\AA})}{10^{42}~{\rm erg~s^{-1}}}\right]^{-0.2}
\end{equation}

The adopted equation for the [O III] case, from \citet{2017MNRAS.468..620Z}, is:

\begin{equation}
\log\left[\frac{\lambda L_{\lambda}(5100\text{\AA})}{{\rm erg~s^{-1}}}\right] = (17.28 \pm 0.98)\cdot \log\left(\frac{L_{\rm [O~III]_{\rm core}}}{{\rm erg~s^{-1}}}\right) + (0.65 \pm 0.02)
\end{equation}

\noindent where $L_{\rm [O~III]_{\rm core}}$ is the luminosity of the core component of the [O III]$\lambda5007$ line. As for the black hole mass, Eddington ratios derived from the [O III]$\lambda5007$ line should be considered as upper limits. The results are summarized in Table~\ref{tab_fit}.

\section{Notes on individual objects}\label{sec_notes}
In the following, we provide the emission lines fitting details for each source, along with the explanation for the new classification. The fitting parameters are reported in Table~\ref{tab_fit}, while the spectra are plotted in Figure~\ref{fig_app_plots}. 

\subsection{J0102$+$4214}\label{secJ0102}
The GTC spectrum is significantly affected by noise. The H$\beta$--[O III] complex shows a prominent bump and is partially obscured by telluric absorption, which may result from Fe II emission or host-galaxy contamination. This prevents us from proposing a definitive classification of the AGN. To extract physical parameters, we fitted the Mg II$\lambda2800$ line, as reported in Table~\ref{tab_fit}. A tentative fit of the H$\beta$ line with a single Lorentzian profile yields FWHM(H$\beta$) = (2931 $\pm$ 980) km s$^{-1}$. This value is consistent with the estimate reported by \citet{2012ApJ...748...49S}, FWHM(H$\beta$) $\sim$ 1900 km s$^{-1}$, but does not support a classification as NLS1. The [O III]$\lambda\lambda4959,5007$ lines are severely affected by telluric absorption and buried in the noise. Consequently, we are unable to provide a revised classification for this object and therefore adopt the NLS1 classification reported by \citet{2022Univ....8..587F}.

\subsection{J0224$+$0700}\label{secJ0224}
The GTC spectrum reveals a few oxygen lines ([O II]$\lambda$3727 and [O III]$\lambda\lambda$4959,5007), as well as Mg II$\lambda$2800 emission. In this case, telluric absorptions are prominent in both the H$\beta$ and H$\gamma$ regions, preventing us from characterizing the line profiles or measuring their FWHM. On the other hand, the Mg II line is relatively unaffected by atmospheric absorption, allowing us to derive the black hole mass and the Eddington ratio. We reported the same classification proposed by \citet{2022Univ....8..587F} as NLS1.

\subsection{J0422$-$0644}\label{secJ0422}
The UT1 spectrum shows multiple emission lines, both permitted (H$\alpha$, H$\beta$, and H$\gamma$) and forbidden ([O III]$\lambda\lambda\lambda$4363, 4959, 5007). The H$\beta$--[O III] region is characterized by prominent Fe II emission, with $R_{4570} \sim 1.19$. After subtracting the Fe II contribution, we fitted the H$\beta$ line using three Gaussian components (two broad and one narrow), and modeled the [O III] lines with single Gaussian profiles. The resulting total FWHM(H$\beta$)=(1940$\pm$119) km s$^{-1}$, placing the object in the classical NLS1 family. Finally, the [O III]/H$\beta$ flux ratio is approximately 0.22, confirming the relative faintness of the oxygen lines compared to the hydrogen lines. From these parameters we can propose a classification as NLS1.

\subsection{J0442$-$0017}\label{secJ0442}
\citet{2024MNRAS.532.3729L} studied in detail the spectral energy distribution (SED) of this AGN proposing a hybrid nature, between NLS1 and FSRQ. The UT1 spectrum is noisy and affected by atmospheric absorption features; however, the H$\beta$--[O III] region is only partially impacted. The Fe II emission contributes significantly, with $R_{4570} \sim 1.05$, although the He II$\lambda$4686 line is also present in the same spectral region. The best fit for the H$\beta$ line was obtained using a Lorentzian profile, while the [O III]$\lambda\lambda$4959,5007 lines were modeled with single Gaussians. The resulting total FWHM(H$\beta$)=(3518$\pm$1218) km s$^{-1}$, outside the range typically associated with NLS1s, although the large uncertainty -- mainly due to the noise level in the H$\beta$ region -- limits the robustness of this measurement. The [O III]/H$\beta$ flux ratio is $\sim$0.90, consistent with type 1 AGN properties. Given the large uncertainty on the FWHM(H$\beta$), our results are consistent with those of \citet{2012ApJ...748...49S} and \citet{2022Univ....8..587F}, while also being compatible with a classification as BLS1. Considering the central value of the H$\beta$ width, we suggest a classification as BLS1, although higher-resolution spectra are required to confirm this result.

\subsection{J0515$-$4556}\label{secJ0515}
The UT1 spectrum reveals all the main optical features, with a not negligible Fe II contribution ($R_{4570} \sim 0.44$). The [O III]$\lambda\lambda$4959,5007 lines display two peculiar characteristics: a blueshift of the entire feature by approximately 132 km s$^{-1}$, and a flux ratio of about 2, which deviates from the theoretical value of $\sim$3 (as can be seen from Figure~\ref{fig_app_plots}). This can be the result of a spurious spike under [O III]$\lambda$4959, or due to the very broad component found in H$\beta$. The H$\beta$ line was modeled with three Gaussian components, and the resulting profile shows a redshifted very broad component ($\Delta v \sim 3700$ km s$^{-1}$). Furthermore, the [O III]/H$\beta$ flux ratio is $\sim$0.77. The asymmetric profile of H$\beta$ can be the result of an NLS1 or and IS with a red wing. Given the $\gamma-$ray detection for this AGN, the orientation favor the type 1 classification for this object, but does not exclude the IS one. The total FWHM(H$\beta$)=(1534$\pm$85) km s$^{-1}$. Due to these considerations, we propose a classification in between NLS1 and IS, to disentangle between the two, intermediate- or high-resolution spectra are needed.

\subsection{J0521$-$1734}\label{secJ0521}
The UT1 spectrum shows a relatively noisy red part, but the H$\beta$--[O III] region is not affected by these fluctuations. The best fit for H$\beta$ was obtained with a single Gaussian component, with its FWHM tied to that of [O III], yielding FWHM([O III])=(845$\pm$30) km s$^{-1}$. The strength of the [O II]$\lambda$3727 line may suggest a possible LINER nature for this AGN. However, a classification as a Seyfert 2 galaxy (Sy2/SY2) appears more likely, given the profiles of [O III]$\lambda$5007 and H$\beta$. In this scenario, J0521$-$1734 could be interpreted either as a misaligned AGN or as a true Sy2 lacking a BLR (e.g., \citealt{2025A&A...695A..55M}). Distinguishing between the LINER and Sy2 scenarios requires high- or intermediate-resolution spectroscopy with broader spectral coverage, ideally including H$\alpha$ to enable the construction of diagnostic diagrams such as those developed by \citet{1981PASP...93....5B} and \citet{1987ApJS...63..295V} (BPT/VO). Moreover, to confirm a possible absence of the BLR, polarized-light observations would be necessary.

\subsection{J0932$+$5306}\label{secJ0932}
In the GTC spectrum, all the main optical features are clearly detected. The iron emission is negligible, which leave space to the detection of the He II$\lambda$4686 line. The best fit for the H$\beta$ line was obtained using a combination of Gaussian components, yielding a total FWHM(H$\beta$)=(1439$\pm$88) km s$^{-1}$. Each of the [O III] lines was modeled with a single Gaussian. The [O III]/H$\beta$ flux ratio is $\sim$1.50, a value consistent with those typically observed in NLS1 galaxies. The relatively narrow H$\beta$ line width and the weakness of the [O III] lines both support a classification as NLS1, despite the absence of significant Fe II emission. Similar cases of NLS1s with faint or undetectable Fe II features have been reported in the literature (e.g., \citealt{2011nlsg.confE...2P}). We therefore confirm the classification as NLS1 proposed by \citet{2022Univ....8..587F}.

\subsection{J1048$-$1912}\label{secJ1048}
The UT1 spectrum shows fluctuations in the H$\beta$--[O III] region. We fitted oxygen lines with a single Gaussian each and a combination of two Gaussians for H$\beta$. The resulting H$\beta$ profile is typical of an IS object, where the narrow and the broad components are clearly separable. The narrow component present a FWHM(H$\beta$) totally dominated by the instrumental broadening, as the one of [O III]$\lambda$5007. The H$\beta$ broad component present a FWHM(H$\beta$)=(6040$\pm$189) km s$^{-1}$ with a significant shift to the red with $\Delta v\sim 1528$ km s$^{-1}$. The global FWHM(H$\beta$) is consistent with the value reported by \citet{1984PASA....5..341M}. The iron contribution is negligible ($R_{4570} \sim 0$), and the [O III]/H$\beta$ ratio is approximately 0.81. This allows us to reclassify the object as an IS, given the composite profile of H$\beta$. Following the classification scheme proposed by \citet{1992ApJS...79...49W}, the intermediate Seyfert type can be obtained calculating the ratio between the fluxes of $R_{Sy-type}$=F([O III])/F(H$\beta$). In this case we obtain a Sy1.5 type. 

\subsection{J1102$+$5251}\label{secJ1102}
The GTC spectrum shows a negligible iron contribution ($R_{4570} \sim 0$) and a peculiar H$\beta$ profile, composed of a narrow core and a prominent red wing. The FWHM of the [O III] lines and of the narrow component of H$\beta$ are entirely dominated by instrumental broadening, whereas the broad component of H$\beta$ has FWHM=(3051$\pm$296) km s$^{-1}$ and is redshifted by $\Delta v \sim 590$ km s$^{-1}$. This line shape can be interpreted as consistent with an IS. The [O III]/H$\beta$ flux ratio is $\sim$1.32, which is compatible with values observed in both classes. Based on these considerations -- particularly the limitations imposed by instrumental broadening -- we propose a classification as IS for this AGN. Further observations with higher spectral resolution are required to confirm this scenario.

\subsection{J1154$+$4037}\label{secJ1154}
The GTC spectrum is affected by several spurious peaks, which were subsequently removed. The final spectrum shows a prominent Mg II$\lambda$2800 line with FWHM=(10098$\pm$297) km s$^{-1}$, while the H$\beta$--[O III] region is completely buried in the noise, preventing a definitive classification. \citet{2022Univ....8..587F} reported an NLS1 classification, but given the broadness of the Mg II$\lambda$2800 line, this does not appear reliable. As a consequence of the broad Mg II$\lambda$2800 line, the calculated black hole mass is the highest in the sample, with log($M_{\rm BH}/M_{\odot}$)=(9.32$\pm$0.06). This result highlights the need for further investigation into the nature of the source, given that NLS1s are usually hosted in low-mass galaxies. We therefore propose to classify it as AMB.

\subsection{J1202$-$0528}\label{secJ1202}
The UT1 spectrum displays all the main optical features, including both permitted and forbidden emission lines. The iron contribution is estimated to be $R_{4570} \sim 1.22$. The H$\beta$ line was fitted with a combination of two Gaussian components, resulting in a total FWHM(H$\beta$)=(1445$\pm$57) km s$^{-1}$, while the [O III]$\lambda\lambda$4959,5007 lines were modeled with two Gaussians each (core and outflow components). The [O III]/H$\beta$ ratio is $\sim$0.83. These properties support the classification of this object as an NLS1.

\subsection{J1246$-$2548}\label{secJ1246}
The H$\beta$ line is fitted with two Gaussian components, resulting in a total FWHM(H$\beta$)=(1430$\pm$266) km s$^{-1}$, which lies within the typical range for NLS1s. The broad component appears redshifted by $\Delta v \sim420$ km s$^{-1}$. The [OIII]$\lambda\lambda$4959,5007 lines are prominent, but their FWHM is completely dominated by the instrumental broadening. The flux ratio [O III]/H$\beta \sim 0.41$ favor the classification as NLS1. The iron emission is negligible, with $R_{4570} \sim 0$, but the prominent telluric absorption in the blue part of the iron emission prevented us to fit properly these lines. Overall, we propose a classification as NLS1. 

\subsection{J1310$+$5514}\label{secJ1310}
In the GTC spectrum, the H$\beta$--[O III] region appears very noisy due to its proximity to the grism efficiency drop\footnote{\url{https://www.gtc.iac.es/instruments/osiris/media/OSIRIS-R1000.jpg}}, and is therefore considered as an artifact. For this reason, we excluded it from the final spectrum. The physical parameters were estimated from the Mg II$\lambda2800$ line, fitted with a Lorentzian profile, which yields FWHM=(2271$\pm$38) km s$^{-1}$. New observations are required to establish a robust classification of this source. We therefore maintained the NLS1 classification proposed by \citet{2022Univ....8..587F}.

\subsection{J1331$-$1325}\label{secJ1331}
The UT1 spectrum displays all the main optical emission lines, from [O II]$\lambda$3727 to [S II]$\lambda\lambda$6716,6731. The iron contribution is negligible ($R_{4570} \sim 0$). H$\beta$ presents a peculiar profile with a red bump ($\Delta v \sim 350$ km s$^{-1}$), best fitted with three Gaussian profiles. The resulting total FWHM(H$\beta$)=(575$\pm$94) km s$^{-1}$. The [O III]$\lambda\lambda$4959,5007 lines are fitted with a single Gaussian each, and the [O III]/H$\beta$ ratio is $\sim$2.21. Considering the shape of the H$\beta$ profile, we suggest an IS classification for this AGN.

\subsection{J1818$+$0903}\label{secJ1818}
The UT1 spectrum displays all the main optical features, from [O II]$\lambda$3727 to [O III]$\lambda$5007. The iron emission yields $R_{4570} \sim 0.97$. The H$\beta$ line was fitted using a Lorentzian component, resulting in a FWHM(H$\beta$)=(1418$\pm$335) km s$^{-1}$. The [O III] lines were fitted with a single Gaussian, but their FWHM is dominated by the instrumental broadening. The [O III]/H$\beta$ ratio is $\sim 0.98$. Based on these results, we suggest a classification of this AGN as NLS1.

\subsection{J1902$-$6748}\label{secJ1902}
The UT1 spectrum covers all the main optical features, including H$\alpha$. The iron contribution is present with $R_{4570} \sim 0.79$, and blended also with the He II$\lambda$4686 line. The best fit for the H$\beta$ line was obtained using a combination of three Gaussian components: the FWHM of the narrow component was tied to that of [O III], while the broad component is composed of two bumps, the broadest of which is redshifted by $\Delta v \sim 2400$ km s$^{-1}$. A similarly asymmetric profile is also visible in the H$\delta$ line. The total FWHM(H$\beta$)=(1015$\pm$117) km s$^{-1}$. The [O III]/H$\beta$ flux ratio is $\sim 2.24$. This case closely resembles the one discussed in Section~\ref{secJ0515}. For this reason, we propose a classification that is intermediate between NLS1 and IS. Further investigations, including higher-resolution spectroscopy, are needed to distinguish between the two scenarios.

\subsection{J2325$-$3559}\label{secJ2325}
The UT1 spectrum displays the typical features of a NLS1: faint [O III]$\lambda\lambda$4959,5007 lines with [O III]/H$\beta \sim 0.58$, strong Fe II emission with $R_{4570} \sim 1.58$, and a global FWHM(H$\beta$)=(971$\pm$75) km s$^{-1}$. The H$\beta$ line was fitted using a single Lorentzian profile, while the [O III] lines show a core component and prominent blue wings with $\Delta v \sim 770$ km s$^{-1}$. Based on these properties, we classify the object as an NLS1.

\subsection{J2354$-$0958}\label{secJ2354}
From the analysis of the UT1 optical spectrum, we conclude that the previous identification of [O II]$\lambda$3727 \citep{2018MNRAS.474.4151D} was incorrect. In the revised interpretation, we identify three lines: Mg II$\lambda$2800, [O II]$\lambda$3727, and H$\gamma$. Based on these features, we calculate a redshift of $z=(0.9920\pm0.0023)$, which is consistent with the value reported by \citet{2015Ap&SS.357...75M}. The spectrum is relatively noisy, and H$\gamma$ lies near the edge of the spectral range, limiting its usefulness for classification purposes. The Mg II line displays a broad profile, pointing toward the BLS1 nature. Due to the absence of significant Balmer lines, we do not propose a new classification for J2354$-$0958 and reported the one proposed by \citet{2022Univ....8..587F} as AMB.

\section{Discussion and conclusions}\label{sec_disc}
In this work, we analyzed the optical spectra of 18 sources extracted from \citet{2022Univ....8..587F}, which in turn are derived from the 4FGL catalog compiled from observations by the {\it Fermi} satellite. The 4FGL catalog is predominantly populated by BL Lac objects, FSRQs, and unclassified sources. However, it also includes other types of AGN, such as misaligned AGN, NLS1s, Seyfert galaxies, and changing-look AGN (CL-AGN) \citep{2022Univ....8..587F}.

\subsection{Classes of $\gamma-$ray emitting AGN}\label{subsec_class}
Our spectroscopic analysis confirms the presence of a variety of AGN types. Specifically, we identified two new NLS1s (J0422$-$0644 and J2325$-$3559), three intermediate Seyferts (J1048$-$1912, J1102$+$5252, and J1331$-$1325), two ambiguous cases showing characteristics between NLS1s and ISs (J0515$-$4556 and J1902$-$6748), one BLS1 (J0442$-$0017), and one Sy2 (J0521$-$1734). In addition, we confirmed the classification of four previously known NLS1s. For the remaining five sources, a reliable classification was not possible due to poor data quality or limited spectral coverage -- as in the case of J2354$-$0958, where the available spectrum was insufficient for a robust analysis. Among these, J1154$+$4037 was reported as ambiguous because the measured width of the Mg II$\lambda$2800 line is no longer compatible with the NLS1 class; however, a new solid classification is still not possible.

\begin{figure}[htbp]
    \centering
    \includegraphics[width=0.5\textwidth]{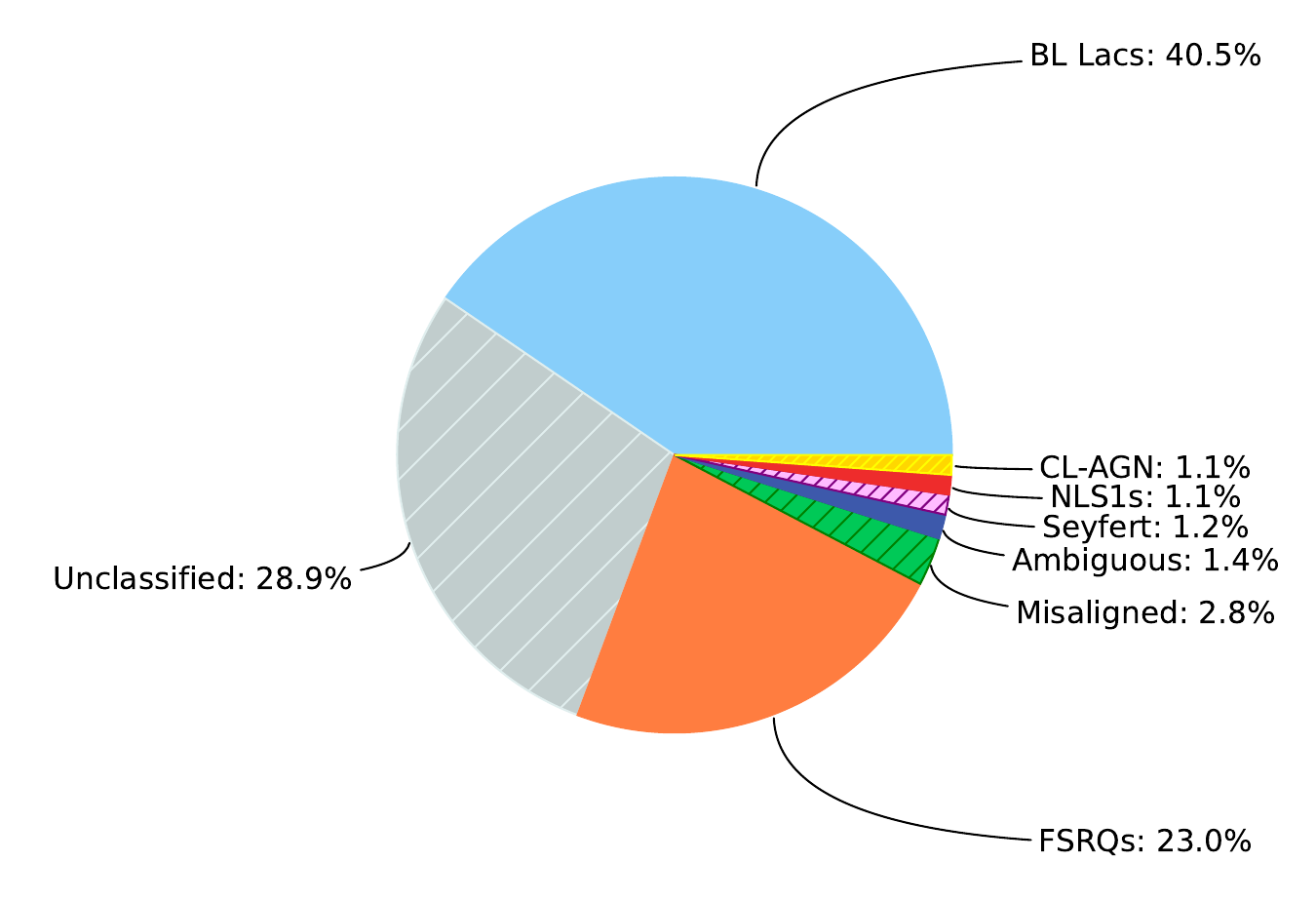}
    \caption{Pie chart showing the distribution of different source classes in the 4FGL catalog. Percentages indicate the updated number of objects in each class, based on \citet{2022Univ....8..587F} and revised according to the analysis presented in this work and other recent studies. Specifically, 2980 sources are from \citet{2022Univ....8..587F}, two (J0105$+$1912 and J0933$-$0013) from \citet{2024MNRAS.527.7055P}, and one (J0959$+$4600) from \citet{2023A&A...676A...9L}.}
     \label{pie}
\end{figure}

When combined with the results of \citet{2022Univ....8..587F}, our findings confirm that different classes of AGN are capable of producing powerful jets detectable in $\gamma$-rays. The separation into classes of jetted AGN, based on the \citet{2022Univ....8..587F} sample together with three additional objects from \citealt{2024MNRAS.527.7055P} and \citealt{2023A&A...676A...9L}, some of which are reclassified according to the analysis presented in this paper, is illustrated in Figure~\ref{pie}. From this plot, it is clear that the majority (about 40\%) are BL Lacs, where the jet dominates the observed spectrum. Another large fraction (about 23\%) consists of FSRQs, $\gamma$-ray sources characterized by strong optical lines and Balmer lines with FWHM $>$ 2000 km s$^{-1}$. A further $\sim$29\% of the sample remains unclassified.  

Beyond these three dominant groups, several minority classes are also represented. The most numerous are misaligned AGN, which include radio galaxies (rdg), soft-spectrum radio quasars (ssrq), and compact steep-spectrum quasars (css). Their classification, based on radio properties and inherited from 4FGL and \citet{2022Univ....8..587F}, remains unchanged in this work. 

Other small fractions include ambiguous objects (1.4\%), Seyferts (1.2\%), NLS1s and CL-AGN (1.1\% both). The presence of $\gamma$-NLS1s remains of particular interest, as it challenges the standard blazar sequence scenario. Their number is steadily increasing, as confirmed by this study. Seyfert-type sources, on the other hand, comprise Sy1, Sy2, IS, and LINERs. While Sy1 objects are compatible with the observed $\gamma$-ray emission, Sy2 (e.g., J0521$-$1734) and IS (e.g., J1048$-$1912, J1102$+$5251, and J1331$-$1325) imply an orientation inconsistent with the jet properties. This misalignment could be due to different mechanisms, such as a recent merger \citep{2020MNRAS.492.1450O}, jet instabilities causing a twist \citep{2024ApJ...964...79L}, or obscuration by interstellar dust in the host galaxy. In the latter case, the AGN may be misaligned with the galaxy disk (e.g., \citealt{2002ApJ...575..150S,2012MNRAS.420..320H}), producing a partial covering of the BLR, this misalignment is usually a consequence of a recent merger. Alternatively, dust could have been redistributed by jet-ISM interaction, introducing line-of-sight obscuration. A variable line-of-sight obscuration could also explain the CL-AGN phenomena (see \citealt{2023NatAs...7.1282R} for a recent review). Finally, in J0521$-$1734, a true Sy2 scenario cannot be excluded. In this case, the BLR is intrinsically absent, so that the AGN is observed face-on but does not display broad permitted lines.

\subsection{Physical parameters}\label{subsec_phys}
The physical parameters derived for our sample span a wide range. The black hole masses, derived by relations that take into account the jet contribution to the continuum optical emission, lie between $\log(M_{\rm BH}/M_{\odot}) = (6.25 \pm 0.40)$ and $(9.32 \pm 0.06)$, with upper limits reaching up to $\log(M_{\rm BH}/M_{\odot}) = 10.23$. The Eddington ratios, instead, range from $R_{\rm Edd} = (2.07 \pm 0.16)$ to $(0.05 \pm 0.03)$.

As expected, both the newly identified and the confirmed NLS1s exhibit black hole masses on the lower end of the AGN spectrum -- typically within the range $10^{6-8}$ $M_{\odot}$, compared to $10^{8-9}$ $M_{\odot}$ for other jetted AGN. In our sample, the NLS1s show a mean value of $\log(M_{\rm BH}/M_{\odot}) = (7.36 \pm 0.68)$. If we include the ambiguous objects with properties between NLS1s and ISs, the mean becomes $\log(M_{\rm BH}/M_{\odot}) = (7.33 \pm 0.62)$, which still falls within the typical NLS1 mass range.

The ISs also exhibit black hole masses within this range. However, it is important to note that in the ISs case the classical virial mass estimates are affected by partial obscuration of the BLR. Since the black hole mass is computed from the broad component of the H$\beta$ emission line flux (see the $L_{\rm H\beta_{broad}}$ term in Equation~\ref{eq_Mbh_sigma}), any attenuation of this component can lead to an underestimation. Therefore, for these sources, the derived black hole mass should be considered a lower limit to the actual value.

The Eddington ratio was calculated by accounting for the possible contribution of the relativistic jet to the optical continuum, particularly when using H$\beta$- and Mg II$\lambda$2800-based estimates. The highest $R_{\rm Edd}$ in our sample are observed in J0224$+$0700 and J2325$-$3559, as expected for NLS1s. NLS1s are typically associated with slim accretion discs, which are thought to be sustained by high, or even super-Eddington, accretion rates \citep{1973A&A....24..337S,1980ApJ...242..772A,1988ApJ...332..646A,1991ApJ...376..214B,1999ApJ...522..839W,2000PASJ...52..499M,2014MNRAS.439..503S}. These regimes are often interpreted as indicative of rapidly growing black holes in AGN at an early evolutionary stage. 

\subsection{The role of the $R_{4570}$ index}\label{subsec_r_fe}
A key diagnostic parameter in our analysis is the $R_{4570}$ index, which quantifies the strength of the optical Fe II emission relative to the H$\beta$ line. This parameter plays a fundamental role in the classification of AGN along the so-called quasar main sequence (MS). The MS represents the distribution of type 1 quasars in the FWHM(H$\beta_{\rm broad}$)--$R_{4570}$ plane, and is widely used to investigate the diversity of quasar spectral properties. Within this plane (see the example in Figure~\ref{MS}), a grid of spectral types has been defined: types A1 through A4 correspond to increasing Fe II emission, whereas B1, B1+, and B1++ indicate increasing FWHM of the broad H$\beta$ component. A horizontal threshold at FWHM(H$\beta_{\rm broad}$) $\sim 4000$ km s$^{-1}$ separates Population A (pop. A) from Population B (pop. B). These populations are believed to group sources sharing broadly similar physical conditions and BLR geometry.

\citet{1992ApJS...80..109B} identified several physical drivers that can influence the position of a quasar within the MS, including: R$_{\rm Edd}$, $M_{\rm BH}$, the BLR cloud covering factor, the anisotropy of the continuum emission, the viewing angle of the source with respect to the observer, the velocity distribution of BLR clouds, and the ionization parameter. These factors contribute to shaping both the line widths and the strength of Fe II emission observed in quasar spectra, and hence their location along the MS. The $R_{4570}$ index therefore serves not only as a classification tool, but also as a tracer of the physical conditions in the nuclear region of AGN \citep{2018FrASS...5....6M}.

In this framework, NLS1s occupy the lower-right region of the MS diagram, due to their relatively narrow H$\beta$ lines and typically strong Fe II emission. In other words, NLS1s belong to pop. A, or even to the so-called extreme pop. A (defined as sources with $R_{4570} > 1$). As described by \citet{2018FrASS...5....6M}, the differences between pop. A and B extend beyond line widths: pop. A sources typically show symmetric, often Lorentzian-shaped Balmer profiles, while pop. B sources exhibit broader, multi-component profiles with possible redward asymmetries. Additionally, lines such as C IV$\lambda$1549, Mg II$\lambda$2800, and the [O III]$\lambda\lambda$4959,5007 doublet often show significant blueshifts in pop. A sources. Other trends include a higher probability of radio-loudness and a flatter soft X-ray slope among pop. B objects \citep{2004AJ....127.1799G,2010MNRAS.403.1759Z}. 

Within our sample, all sources classified as NLS1 exhibit symmetric H$\beta$ profiles, fitted either with a single Lorentzian or multiple Gaussians. Asymmetric Balmer lines, instead, are found in sources classified as IS or in the ambiguous cases between NLS1 and IS. All NLS1s in our sample belong to pop. A, as expected by construction. The ambiguous cases also fall within pop. A. One exception is the source J1048$-$1912, which displays a very broad H$\beta_{\rm {broad}}$ line (FWHM $\sim 6101$ km s$^{-1}$) and lacks detectable Fe II emission; it is therefore assigned to pop. B. These considerations strength our classification, within the MS framework. 

\begin{figure}[htbp]
    \centering
    \includegraphics[width=0.5\textwidth]{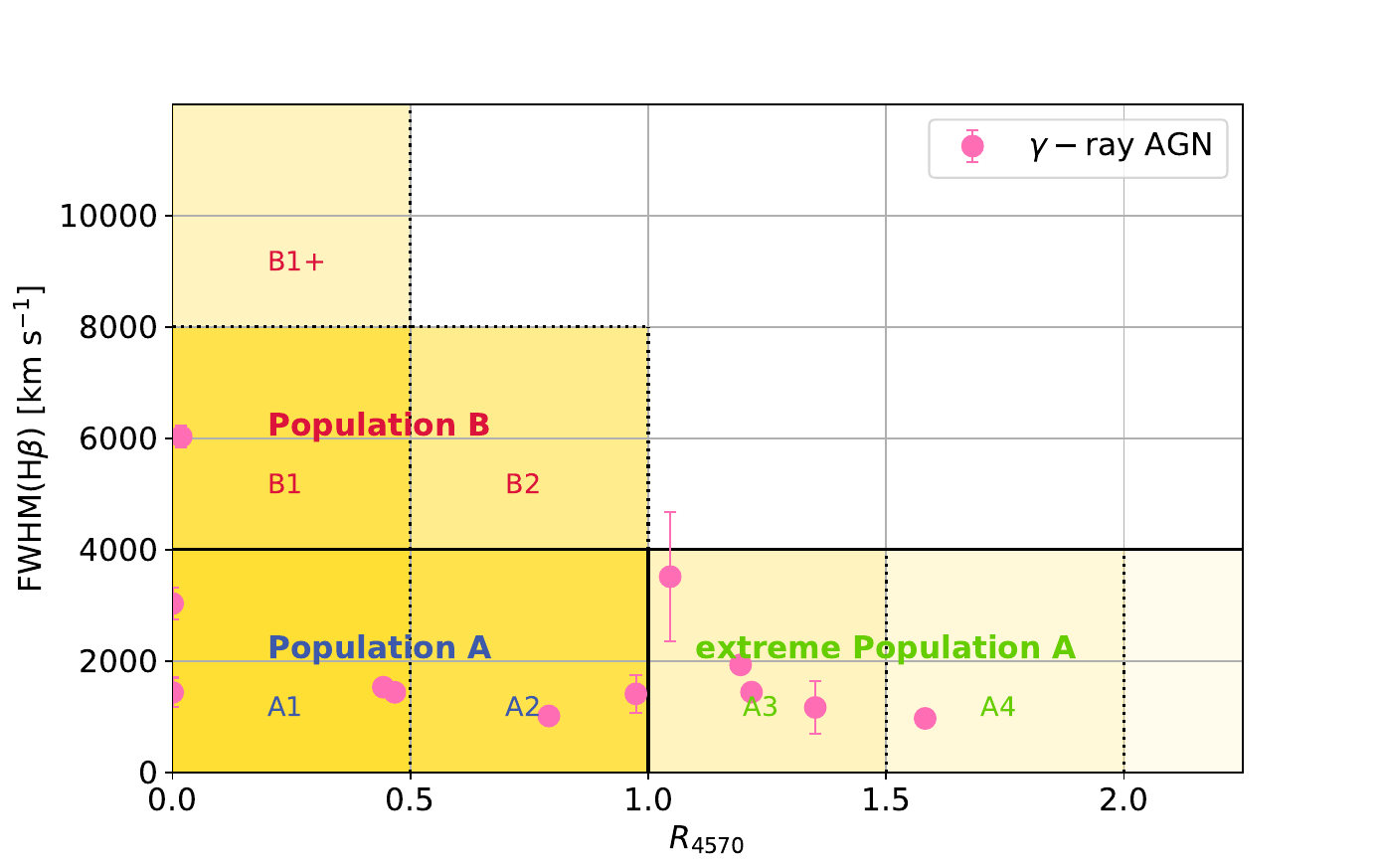}
    \caption{FWHM(H$\beta_{\rm broad}$)--$R_{4570}$ plane defining the quasar MS. The same plot shows pop. A and B, together with their finer subdivisions. Shaded regions indicate the areas populated by quasars, with color intensity representing the source density in each region. The pink points represent the objects in the current sample for which the FWHM(H$\beta_{\rm broad}$) and $R_{4570}$ were measured.}
     \label{MS}
\end{figure}

\subsection{Conclusions}\label{subsec_concl}
We updated the number of optically and spectroscopically confirmed NLS1s with $\gamma-$ray emission to a total of 26 sources. This number may increase to 36 when including ambiguous cases -- 34 listed in Table~\ref{tab_gamma_nls1} and two additional NLS1/IS borderline objects reported in this work. The complete list of $\gamma-$ray-emitting NLS1s is provided in Table~\ref{tab_gamma_nls1}.

To resolve the ambiguous classification and to provide a more robust characterization of those sources for which we currently report only the classifications from \citet{2022Univ....8..587F}, higher spectral resolution datasets (e.g., as obtained by VLT/XShooter) are necessary. The X-Shooter spectra could allow us to perform a detailed optical analysis of the selected sources, enabling accurate measurements of line profiles and physical parameters. They could also facilitate the identification of peculiar cases, similar to the one recently reported by \citet{2025A&A...698A.320D}, and ultimately improve our understanding of the diversity among $\gamma-$ray-emitting AGN.

\begin{acknowledgements} B.D.B. thank Dr. P. Ratnaparkhi and Dr. A. Costa for the for the fruitful discussion during the "Extragalactic jets at all scales: a Cretan view" conference (Heraklion, August 25-29). C.M. acknowledges support from Fondecyt Iniciacion grant 11240336 and the ANID BASAL project FB210003. Based on observations collected at the European Southern Observatory under ESO programms 110.23UC.003 and 111.24P0.002. Based on data obtained from the ESO Science Archive Facility with DOI: https://doi.org/10.18727/archive/77. Based on observations made with the Gran Telescopio Canarias (GTC), installed at the Spanish Observatorio del Roque de los Muchachos of the Instituto de Astrofísica de Canarias, on the island of La Palma. 
\end{acknowledgements}

\bibliographystyle{aa} 
\bibliography{./biblio.bib}

\begin{appendix}\label{appendix}
\onecolumn

\begin{landscape}
\section{Fitting parameters and classification}\label{appendix_fit} 
\justifying

\begin{table}[h!]
\small
\centering
\caption{Fitting parameters with the corresponding uncertainties, classifications, physical parameters, and measured redshift.} 
	\begin{tabular}{l | lll | lll | ll | lll | l}
	   \hline
		 {\bf Name}& {\bf Type} & {\bf Flux}	& {\bf FWHM} &{\bf Type} & {\bf Flux}	& {\bf FWHM}	& {\bf Class. F22}  & {\bf New class.} & {\bf log(M$_{\rm BH}$/M$_{\odot}$)} & {\bf R$_{\rm Edd}$} & {\bf Methods}& {\bf z$_{m}$} \\
				& & {\bf [$10^{-17}$erg s$^{-1}$]} & {\bf [km s$^{-1}$]}  & & {\bf [$10^{-17}$erg s$^{-1}$]} & {\bf [km s$^{-1}$]} & & & & & & \\
           	 \hline
           	\hline
		 	{\bf} & \multicolumn{3}{c}{\bf H$\beta$} & \multicolumn{3}{c}{\bf [O III]$\lambda$5007} & & & & & & \\
	    	\hline
	    	\hline
		J0422$-$0644    		& G$_{\rm n}$  	& 43 $\pm$ 6 	& 572*						& G$_{\rm c}$ & 	58 $\pm$ 3	& 572 $\pm$ 36 & FSRQ & {\bf NLS1} & 7.05 $\pm$ 0.53	& 0.13 $\pm$ 0.07 & (A,D)  & 0.2417 $\pm$ 0.0007 \\
							& G$_{\rm b}$  	& 224 $\pm$ 46 	& 2265 $\pm$ 385  			& -  			& - 			& -   & & &  & & & \\ 	
		\hline
		J0442$-$0017      		& L  		& 120 $\pm$ 14 	& 3518 $\pm$ 1218				& G$_{\rm c}$ & 	81 $\pm$ 5 	& @ & NLS1 & {\bf BLS1} & 7.60 $\pm$ 0.43 & 0.23 $\pm$ 0.10 & (A,D)  & 0.8467 $\pm$ 0.0016  \\
		\hline
		J0515$-$4556    		& G$_{\rm n}$  	& 24 $\pm$ 2 	& 192*						& G$_{\rm c}$ &	 116 $\pm$ 1 	& 192 $\pm$ 6 & AMB & {\bf NLS1/IS} & 7.09 $\pm$ 0.66 & 0.05 $\pm$ 0.03 & (A,D)   & 0.1946 $\pm$ 0.0004 \\
							& G$_{\rm b}$  	& 126 $\pm$ 5 	& 1807 $\pm$ 291  				& -  			& - 			& -   & & &  & & & \\ 
		\hline
		J0521$-$1734		      	& G$_{\rm n}$	& 14 $\pm$ 1 	& 845*						& G$_{\rm c}$ & 	43 $\pm$ 1 	& 845 $\pm$ 30 & FSRQ	 & {\bf SY2} & <10.23 & <0.01 & (B,E)	 & 0.3472 $\pm$ 0.0002	\\
		\hline
		J0932$+$5306       		& G$_{\rm n}$  	& 36 $\pm$ 4  	& 267*						& G$_{\rm c}$ & 	316 $\pm$ 2 	& 267 $\pm$ 5 & NLS1 & {\bf NLS1} & 7.74 $\pm$ 0.50 & 0.11 $\pm$ 0.06 & (A,D)  & 0.5967 $\pm$ 0.0002 	\\
							& G$_{\rm b}$	& 174 $\pm$ 10	& 1813 $\pm$ 359 			& -  			& - 			& -  & & &  & & & \\ 
		\hline
		J1048$-$1912       		& G$_{\rm n}$  	& 19 $\pm$ 1 	& @*		    				& G$_{\rm c}$ & 	115 $\pm$ 1 	& @ & NLS1& {\bf IS} & 8.01 $\pm$ 0.94 & <0.09 & (A,D)  & 0.5953 $\pm$ 0.0002   	\\
							& G$_{\rm b}$  	& 124 $\pm$ 3 	& 6040 $\pm$ 189 				& -  			& - 			& -   & & &  & & & \\       
		\hline 
		J1102$+$5251   		& G$_{\rm n}$  	& 28 $\pm$ 3 	& @*						& G$_{\rm c}$ & 	143 $\pm$ 2 	& @ & NLS1 & {\bf IS} & 7.38 $\pm$ 1.61 & <0.49 & (A,D)	  & 0.6899 $\pm$ 0.0009 \\
							& G$_{\rm b}$  	& 80 $\pm$ 5 	& 3043 $\pm$ 293 	   			& -  			& - 			& -   & & &  & & & \\                 
		\hline
		J1202$-$0528    		& G$_{\rm n}$  	& 174 $\pm$ 18 	& @*					& G$_{\rm c}$ &  	791 $\pm$ 23 	& @ & NLS1 & {\bf NLS1} & 7.86 $\pm$ 0.39 & 0.14 $\pm$ 0.06 & (A,D)  & 0.3805 $\pm$ 0.0002	\\
							& G$_{\rm b}$  	& 954 $\pm$ 80 & 1843 $\pm$ 401   				& G$_{\rm o}$ &	498 $\pm$ 24 	& 1069 $\pm$ 45 & & & & & \\
		\hline
		J1246$-$2548    		& G$_{\rm n}$  	& 86 $\pm$ 7 	& @*						& G$_{\rm c}$ &  	247 $\pm$ 5 	& @  & NLS1 & {\bf NLS1} & 8.22 $\pm$ 0.88 & 0.10 $\pm$ 0.09	& (A,D)  & 0.6372 $\pm$ 0.0002 	\\
							& G$_{\rm b}$  	& 518 $\pm$ 14 & 4779 $\pm$ 149   				& -  			& - 			& -   & & &  & & & \\ 		
	   \hline
		J1331$-$1325    		& G$_{\rm n}$  	& 15 $\pm$ 2 	& @* 						& G$_{\rm c}$ & 	84 $\pm$ 1 	& @ & FSRQ & {\bf IS} & 6.48 $\pm$ 1.55 & <0.20 & (A,D)  & 0.2514 $\pm$ 0.0002	\\
							& G$_{\rm b}$  	& 23 $\pm$ 3 	& 1175 $\pm$ 485    				& -  			& - 			& -   & &  & & & & \\ 	
		\hline         
		J1818$+$0903 			& L  		& 91 $\pm$ 8 	& 1418 $\pm$ 335 					& G$_{\rm c}$ & 	120 $\pm$ 3 	& @ & NLS1 & {\bf NLS1} & 6.25 $\pm$ 0.40 & 0.79 $\pm$ 0.32 & (A,D)  & 0.3542 $\pm$ 0.0004 \\
		\hline              			    
		J1902$-$6748   		& G$_{\rm n}$  & 55 $\pm$ 3 	& @*						& G$_{\rm c}$ & 	432 $\pm$ 2 	& @ & FSRQ  & {\bf NLS1/IS} & 7.21 $\pm$ 0.78 & < 0.11 & (A,D) & 0.2542 $\pm$ 0.0002 	\\
							& G$_{\rm b}$  	& 137 $\pm$ 11 & 1850 $\pm$ 799    			& -  			& - 			& -   & &  & & & & \\ 		
		\hline 
		J2325$-$3559      		& L	 	& 212 $\pm$ 3 	& 971 $\pm$ 75		     			& G$_{\rm c}$ & 	125 $\pm$ 7 	& @ & AMB  & {\bf NLS1} & 6.32 $\pm$ 0.37 & 1.36 $\pm$ 0.51 & (A,D) & 0.3666 $\pm$ 0.0004	\\
							& -  			& - 			& - 							& G$_{\rm o}$ & 	70 $\pm$ 13 	& 1424 $\pm$ 963 & & & & & \\	
	   	\hline  
            	\hline
		 	{\bf } & \multicolumn{3}{c}{\bf Mg II $\lambda$2800} & \multicolumn{3}{c}{\bf [O II]$\lambda$3727} & & & & & & \\
	   	\hline
	    	\hline
		J0102+4214    			& L   	& 65 $\pm$ 1 	& 2771 $\pm$ 82 					& - & - & - &  NLS1 & NLS1 & 7.89 $\pm$ 0.06 & 0.74 $\pm$ 0.01 & (C,F)  & 0.8773 $\pm$ 0.0005	\\
	    	\hline
		J0224$+$0700     		& L  			& 86 $\pm$ 9		& 1408 $\pm$ 229   		& G$_{\rm c}$ &	41 $\pm$ 4 	& @		 & NLS1 & NLS1 & 7.08 $\pm$ 0.34	& 2.07 $\pm$ 0.16 & (C,F)	 & 0.5119 $\pm$ 0.0007 	\\
		\hline					
		J1154$+$4037			& L 			& 223 $\pm$ 3 		& 10098 $\pm$ 297		& G$_{\rm c}$ & 4 $\pm$ 1 	& <454 & NLS1	& {\bf AMB} & 9.32 $\pm$ 0.06 & 0.08 $\pm$ 0.01 & (C,F)  & 0.9251 $\pm$ 0.0004 	\\
	   	 \hline	
	    	J1310$+$5514     		& L 			& 97 $\pm$ 1 		& 2271 $\pm$ 38 		& - & - & - & NLS1 & NLS1 & 7.84 $\pm$ 0.04 & 1.28 $\pm$ 0.01 & (C,F)  & 0.9251 $\pm$ 0.0004 \\ 
	   	\hline				
            	J2354$-$0958			& L 			& 42.5 $\pm$ 0.9 	& 4540 $\pm$ 158 		& G$_{\rm c}$ & 6.0 $\pm$ 0.3 	& 541 $\pm$ 56 & AMB & AMB & 7.62 $\pm$ 0.07 & 0.10 $\pm$ 0.01 & (C,F)   & 0.9884 $\pm$ 0.0005 \\
            \hline
	\end{tabular}
	\tablefoot{Columns: name of the object, type of curve used to fit the components of the line (Lorentzian - L, narrow Gaussian - G$_{\rm n}$, broad Gaussian - G$_{\rm b}$, core Gaussian - G$_{\rm c}$, and outflow Gaussian - G$_{\rm o}$), flux, FWHM of the line along with the uncertainty, old classification from \citet{2022Univ....8..587F} (F22), the new proposed classification, $M_{BH}$, $R_{Edd}$, the used methods, and measured redshift. The FWHM values marked with * correspond to those associated with the narrow lines, while values marked with @ correspond to FWHM totally dominated by the instrumental contribution. The table is divided horizontally into two sections: the H$\beta$--[O III]$\lambda$5007 section and the Mg II$\lambda$2800--[O II]$\lambda$3727 section. The classifications highlighted in bold correspond to cases in which the new spectra were useful either to confirm the classification proposed by \citet{2022Univ....8..587F} or to propose a new one. Methods: (A) for \citet{2020ApJ...903..112D}, (B) for \citet{2014ApJ...789...17H}, and (C) for \citet{2012MNRAS.427.3081T}. The same is valid for the Eddington ratio, due its dependence on $M_{BH}$. In this case the methods are: (D) for the classical derivation of $R_{Edd}$ from H$\beta$, (E) for the derivation of $R_{Edd}$ from [O III]$\lambda$5007, and (F) for the derivation of $R_{Edd}$ from Mg II$\lambda$2800.}
	\label{tab_fit}
\end{table}
\end{landscape}

\FloatBarrier

\section{Plots}\label{appendix_plots}
\begin{figure*}[htbp]
    \centering
    \includegraphics[width=0.44\textwidth]{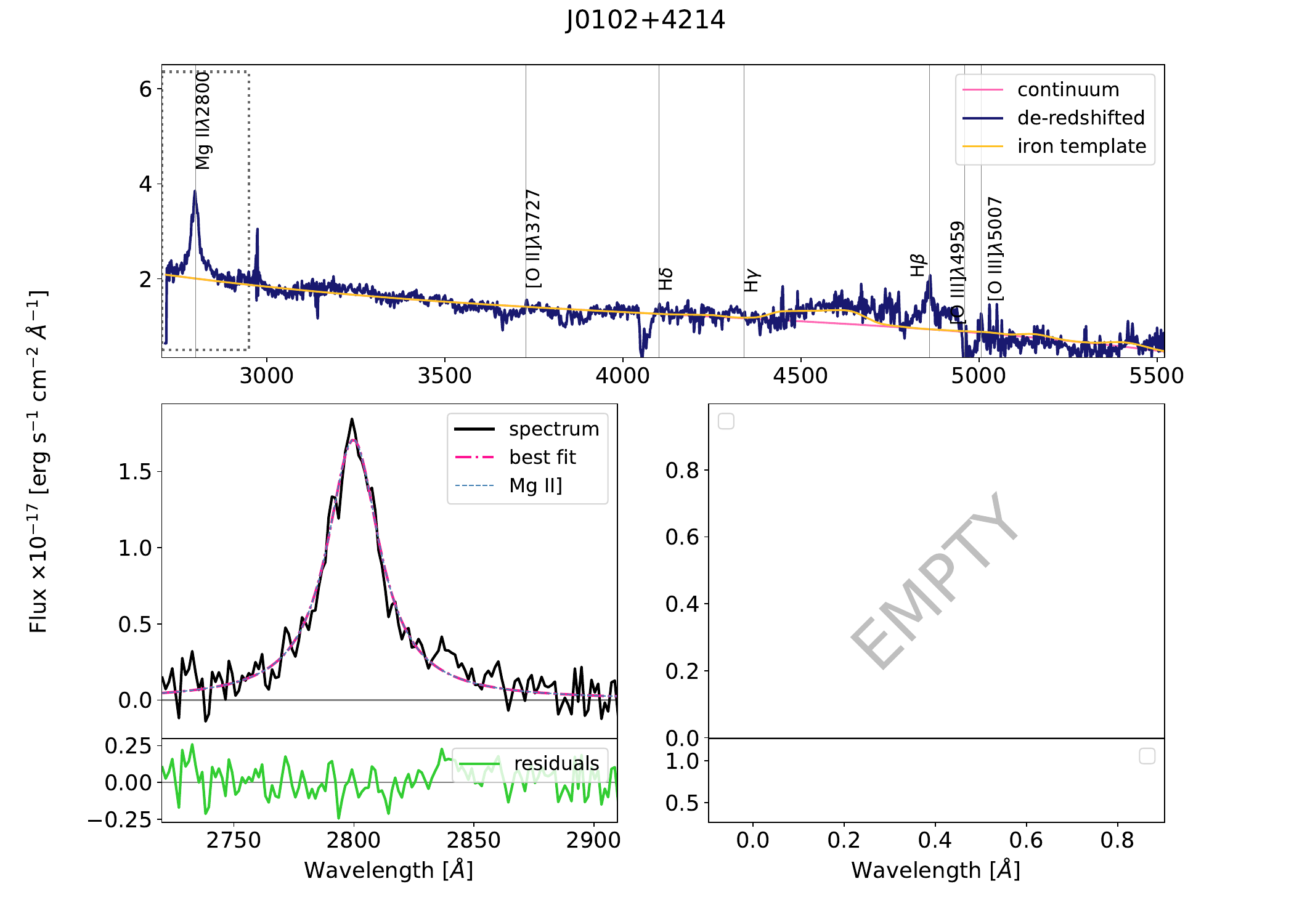}
    \includegraphics[width=0.44\textwidth]{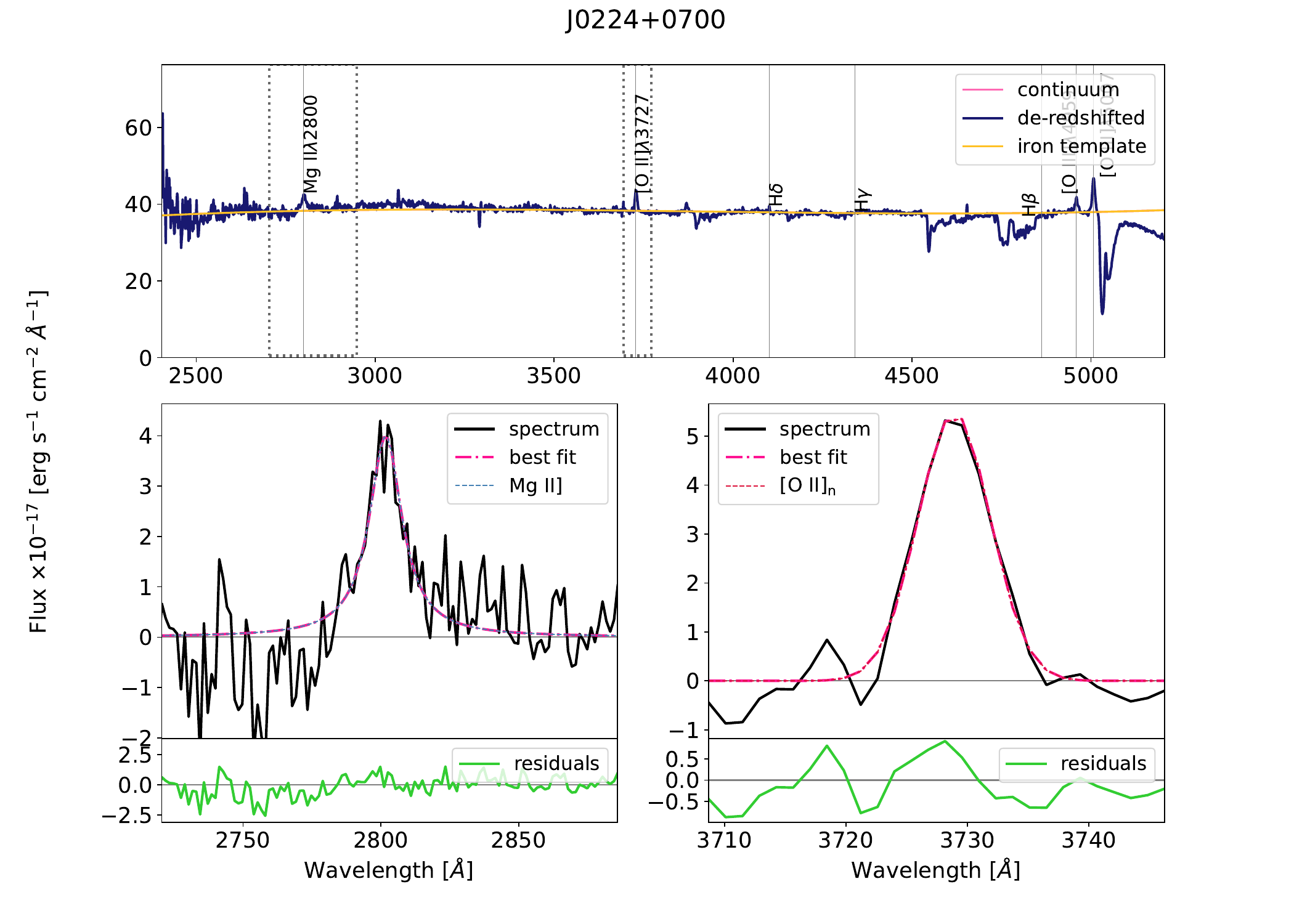}
    \includegraphics[width=0.44\textwidth]{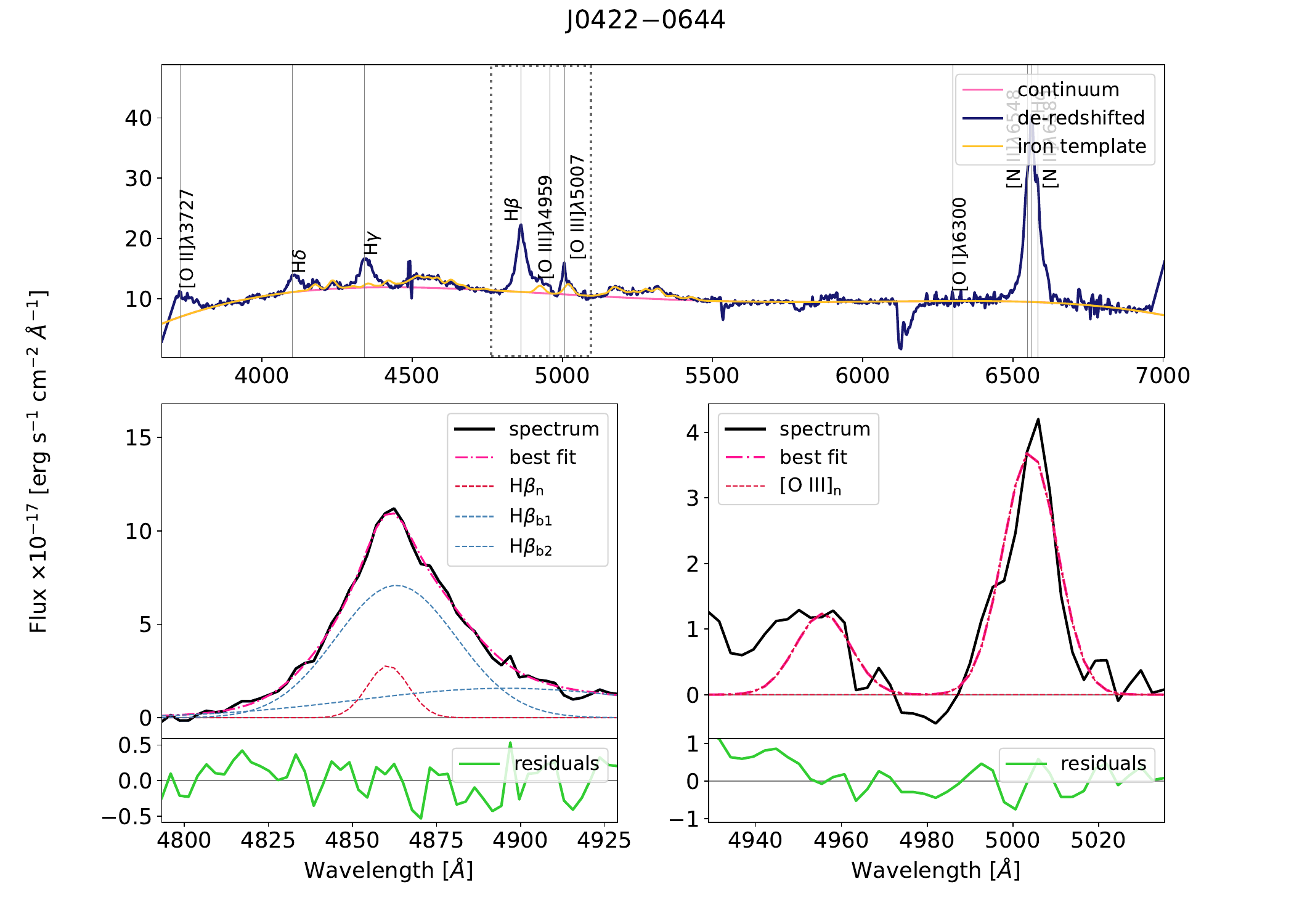}
    \includegraphics[width=0.44\textwidth]{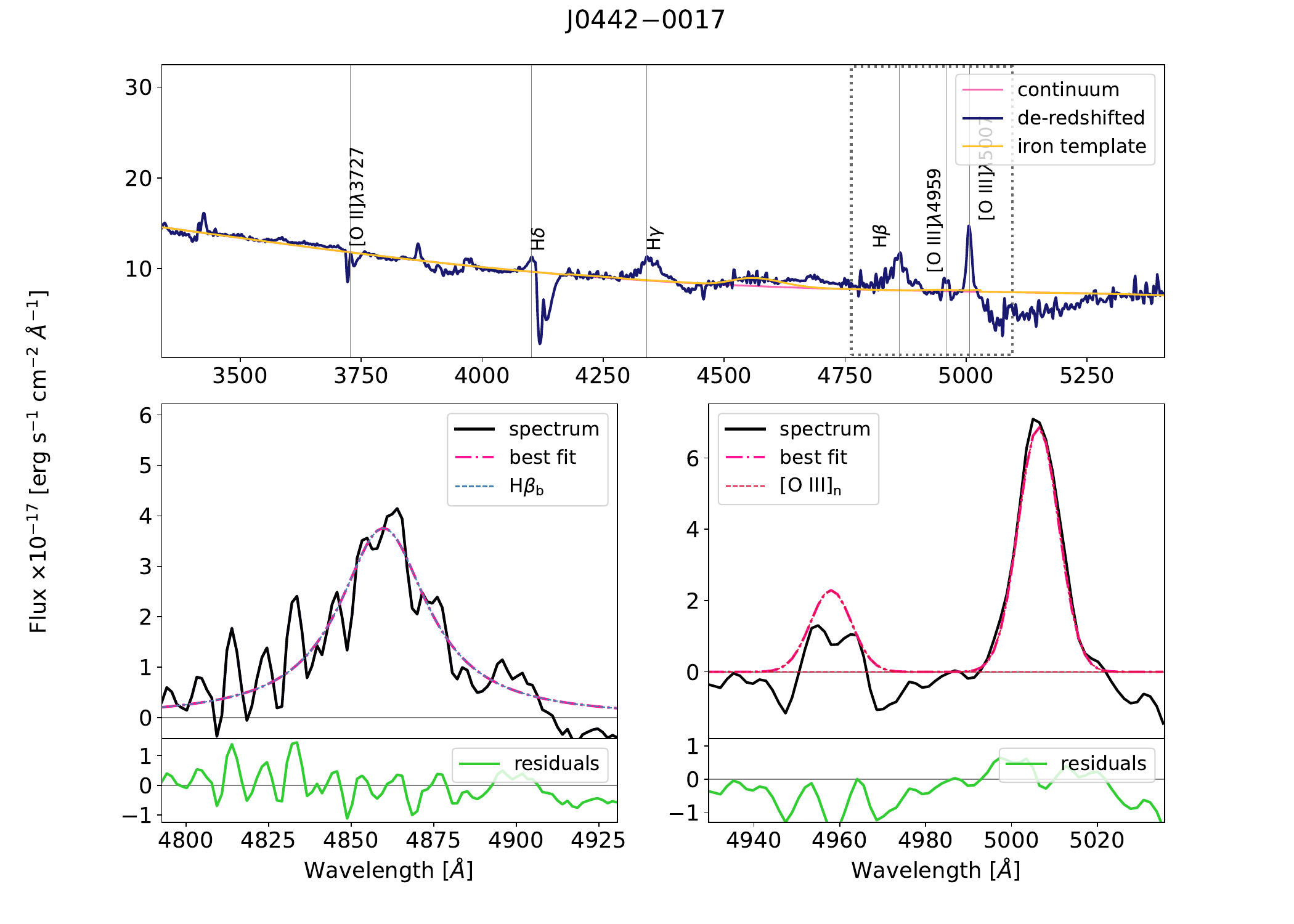}
    \includegraphics[width=0.44\textwidth]{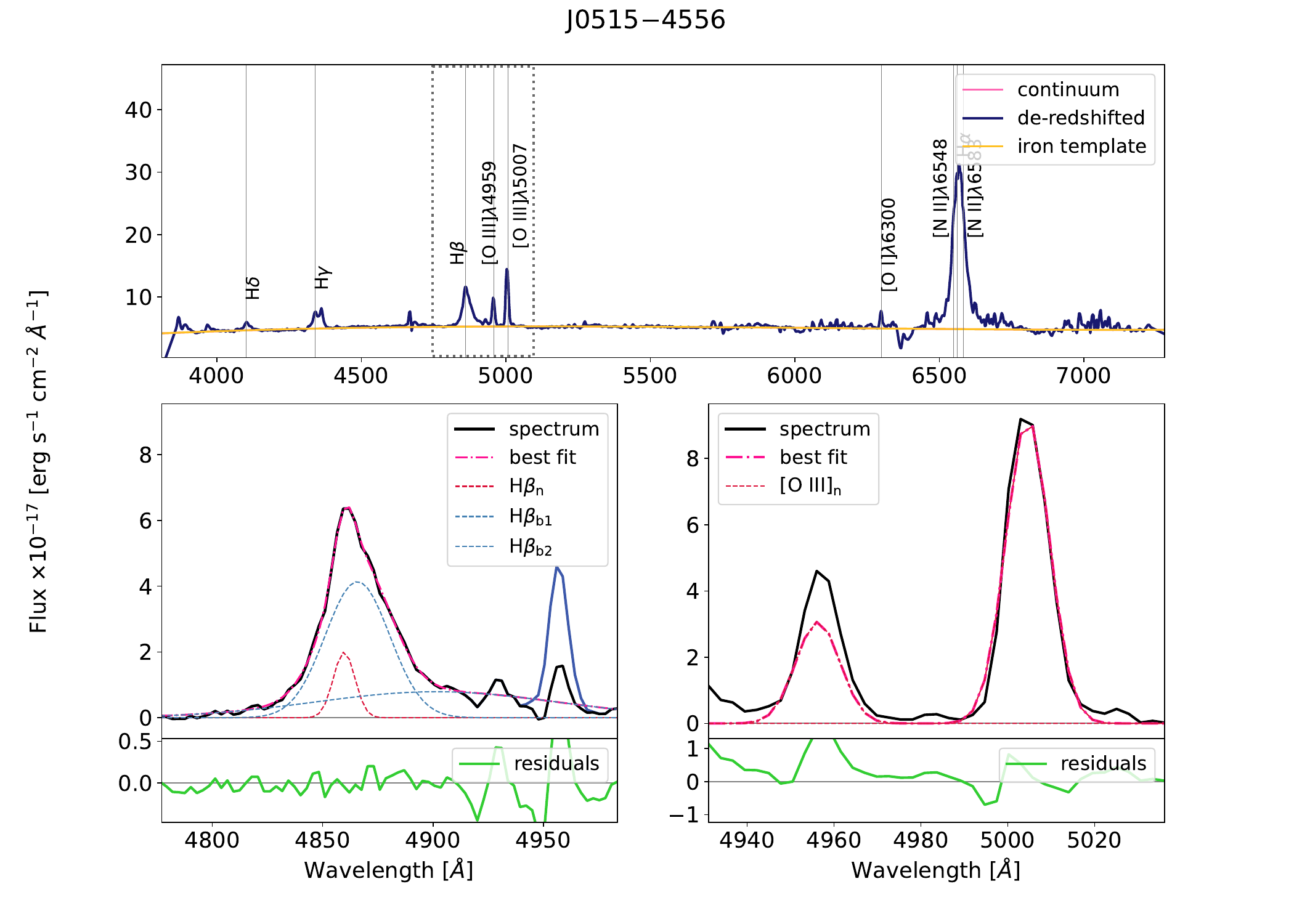}
    \includegraphics[width=0.44\textwidth]{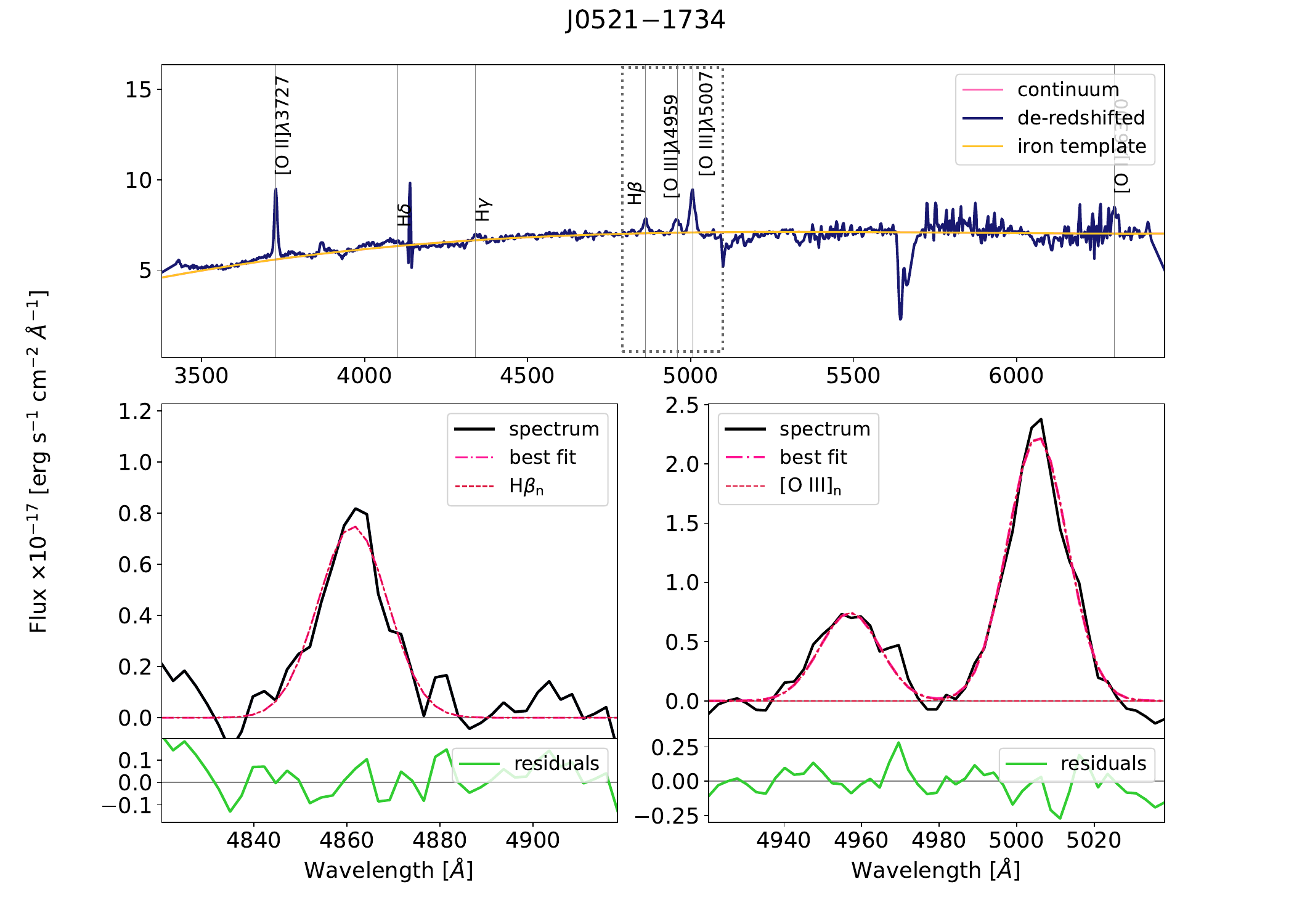}
    \caption{Continued on the next page. Spectra of the considered sources. Top panels: redshift-corrected spectra with continuum and iron line fits overlaid. Grey dotted rectangles indicate the regions that are magnified in the lower panels, where detailed line fitting is performed. Bottom panels: fitted emission lines (e.g., H$\beta$--[O III]$\lambda\lambda$4959,5007 or Mg II$\lambda2800$--[O II]$\lambda3727$), along with the corresponding residuals.}
     \label{fig_app_plots}
\end{figure*}
\begin{figure*}
    \centering 
     \includegraphics[width=0.44\textwidth]{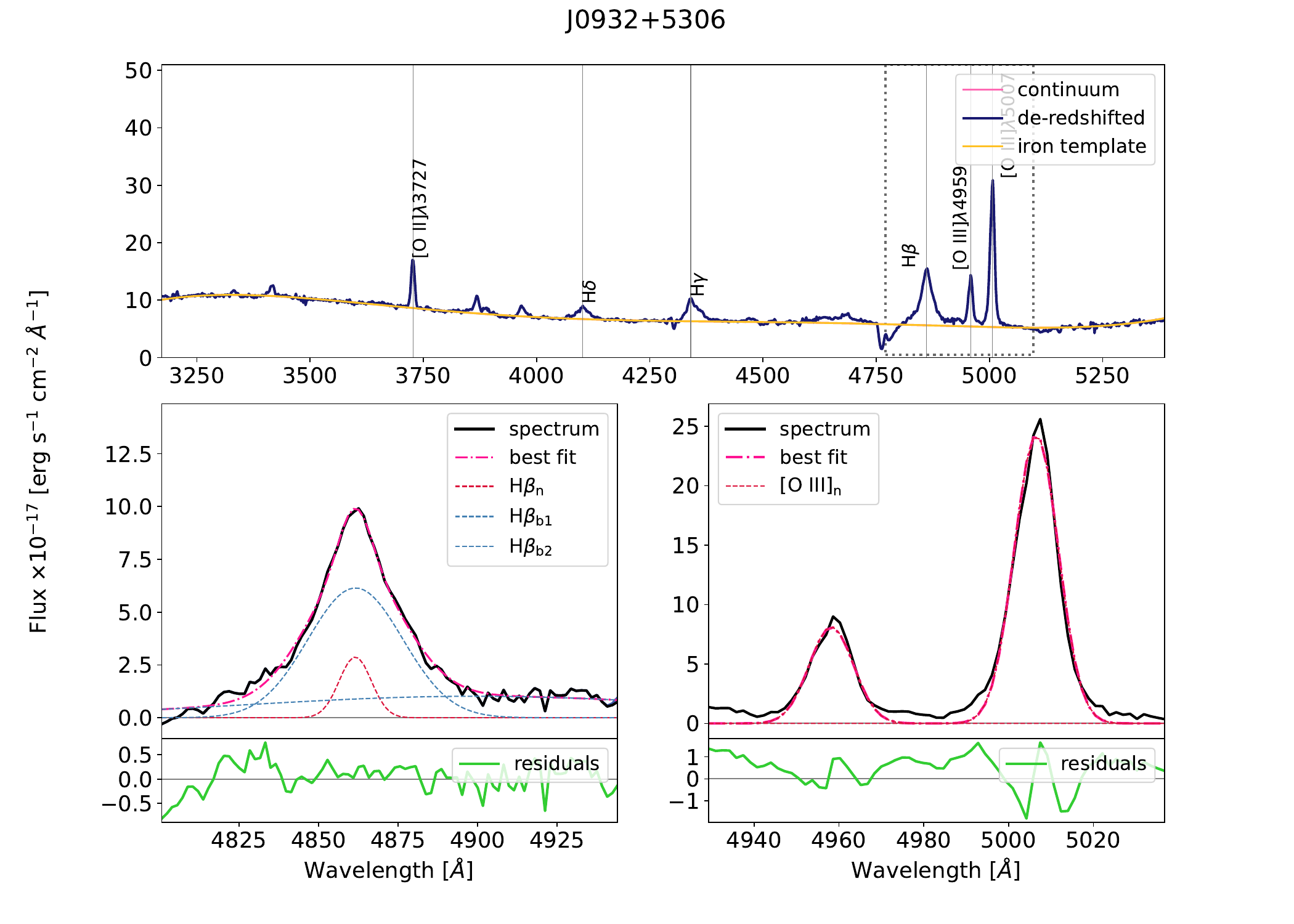}
     \includegraphics[width=0.44\textwidth]{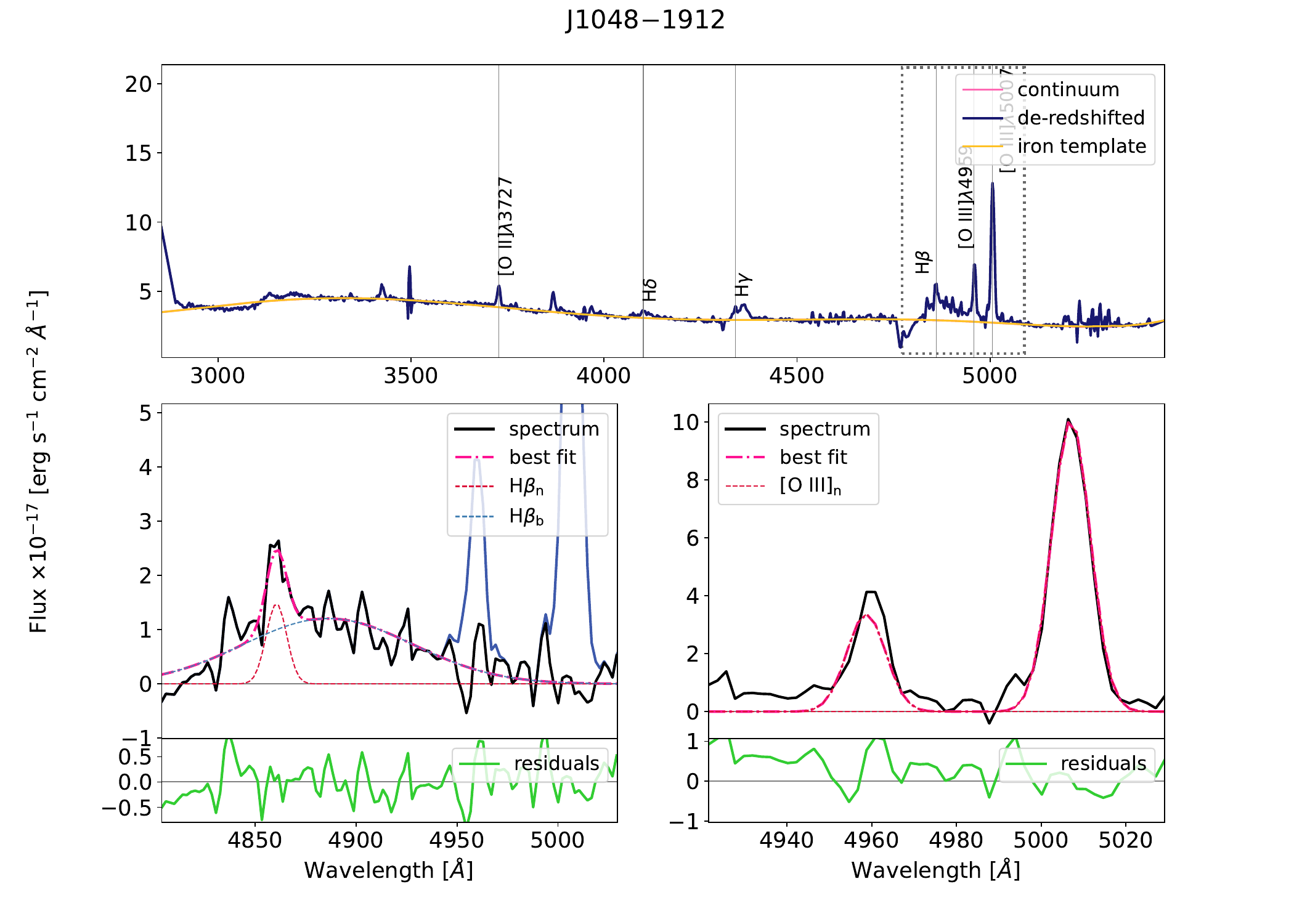}

    \includegraphics[width=0.44\textwidth]{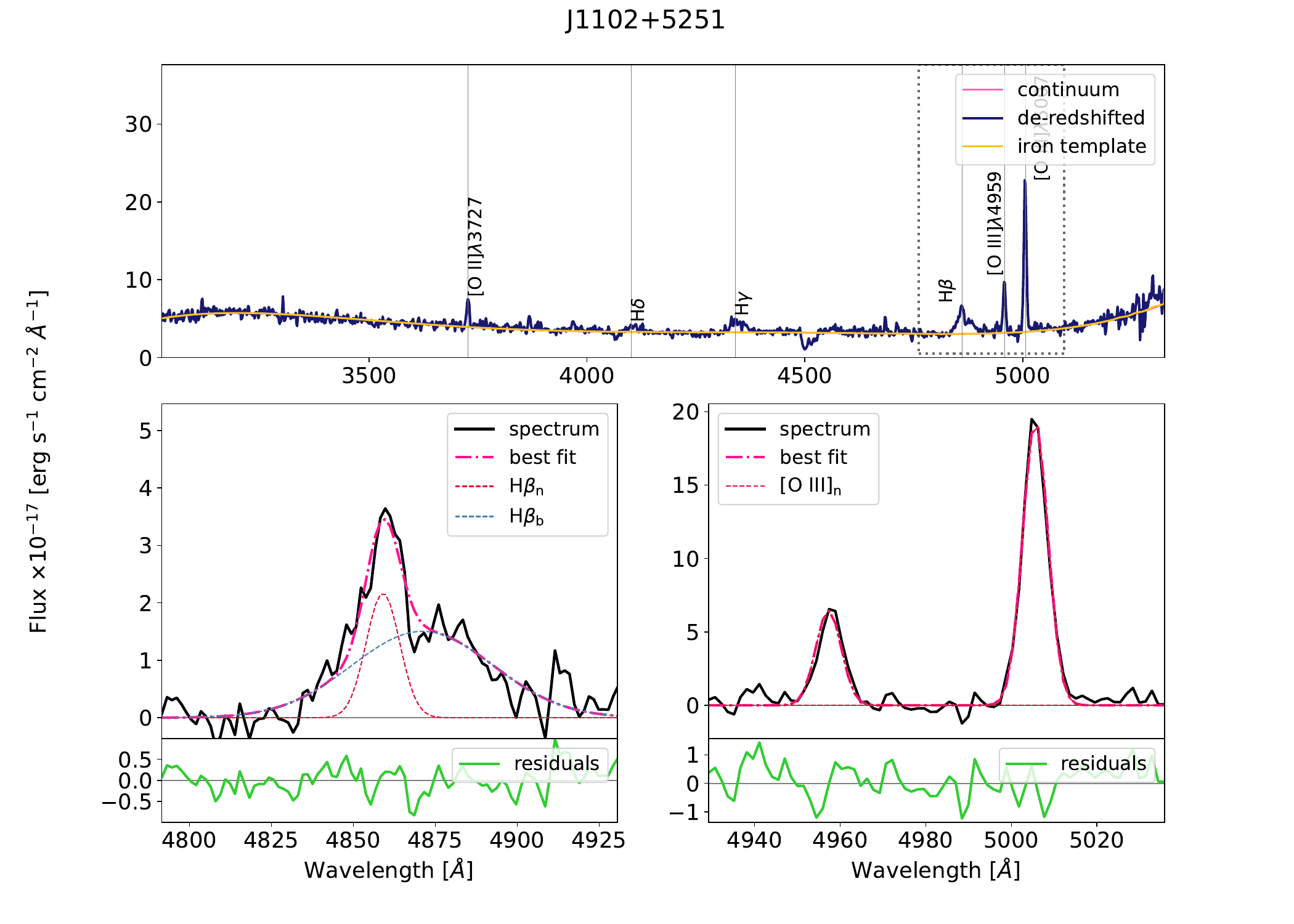} 
    \includegraphics[width=0.44\textwidth]{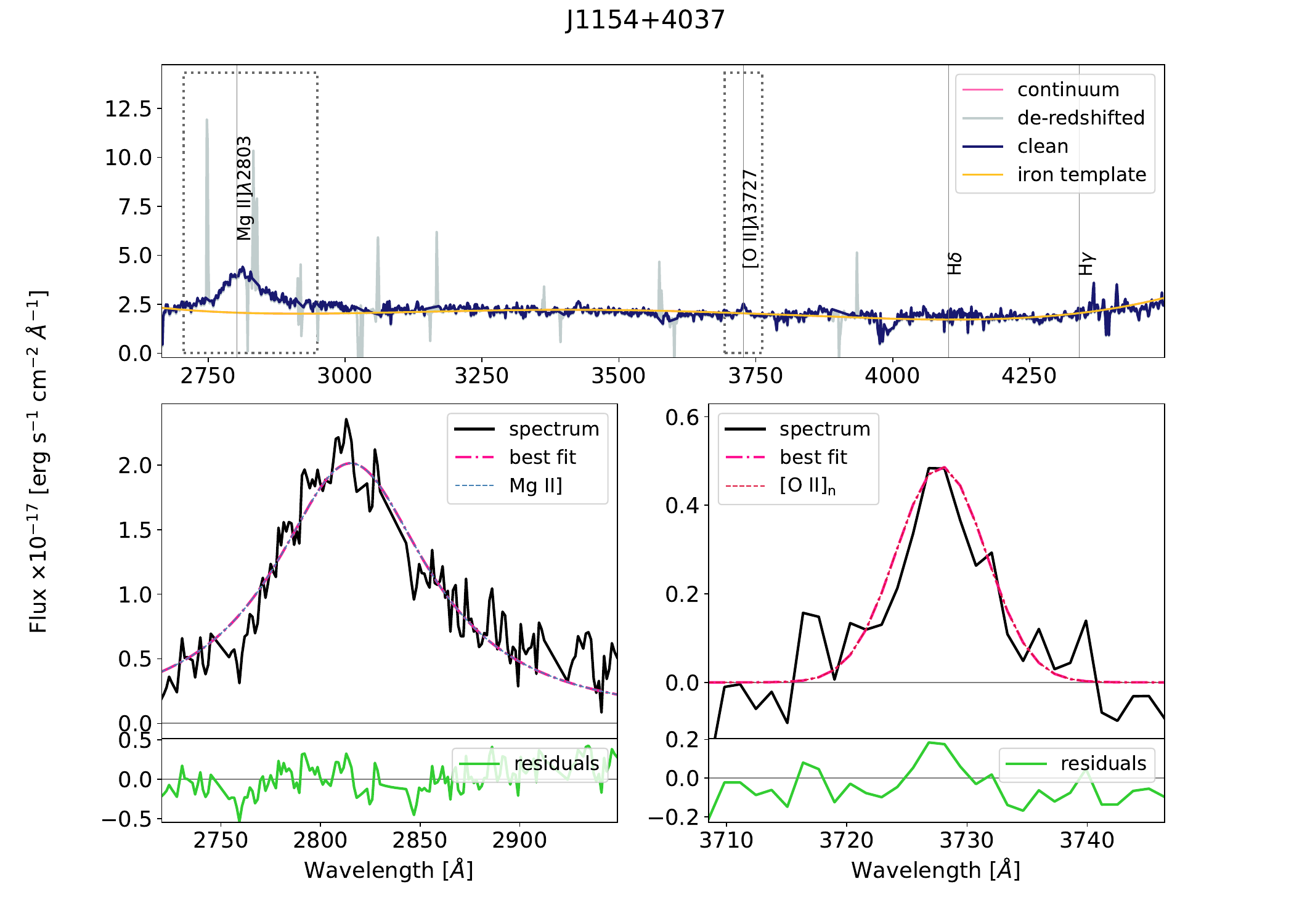} 
     \includegraphics[width=0.44\textwidth]{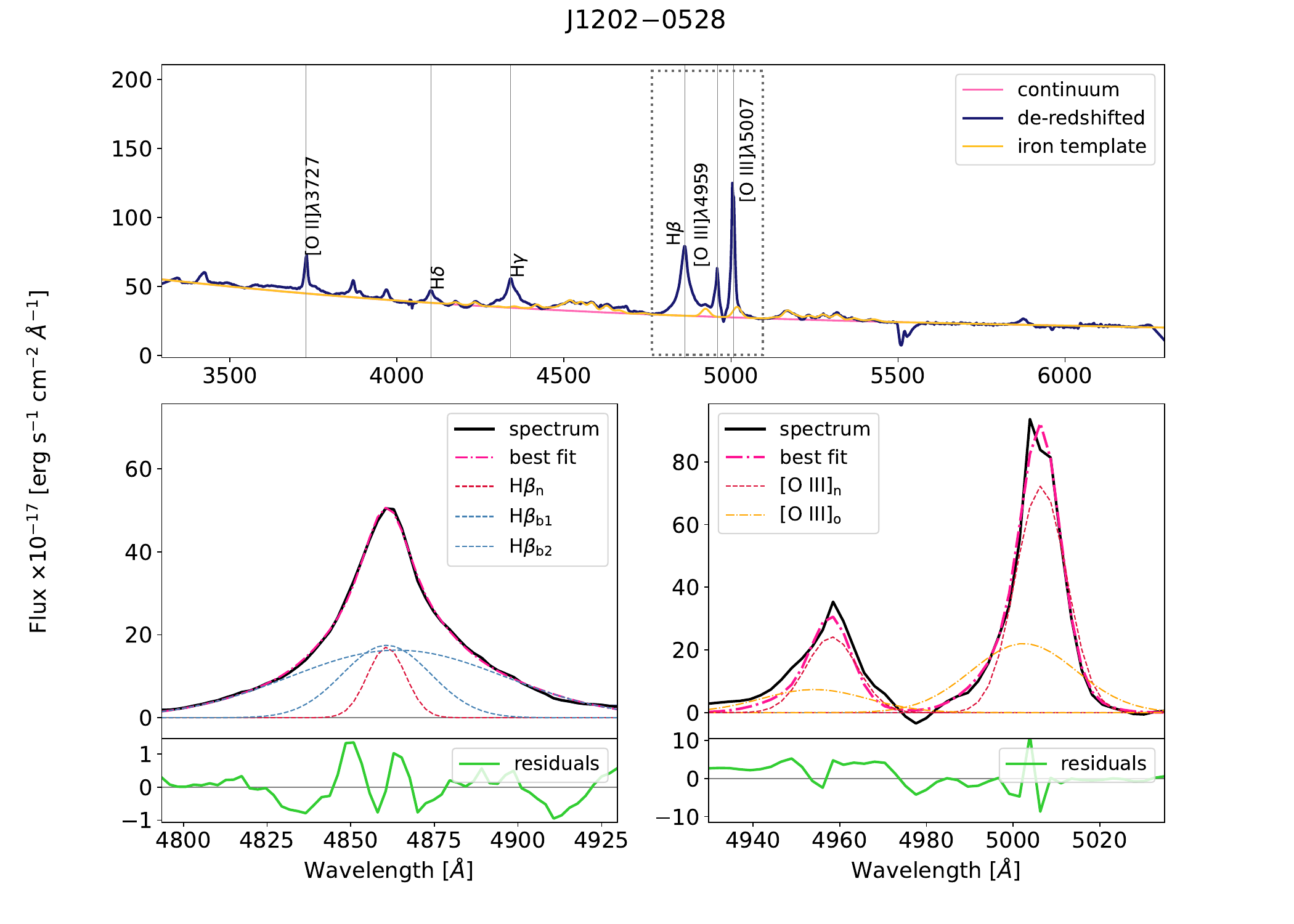}
     \includegraphics[width=0.44\textwidth]{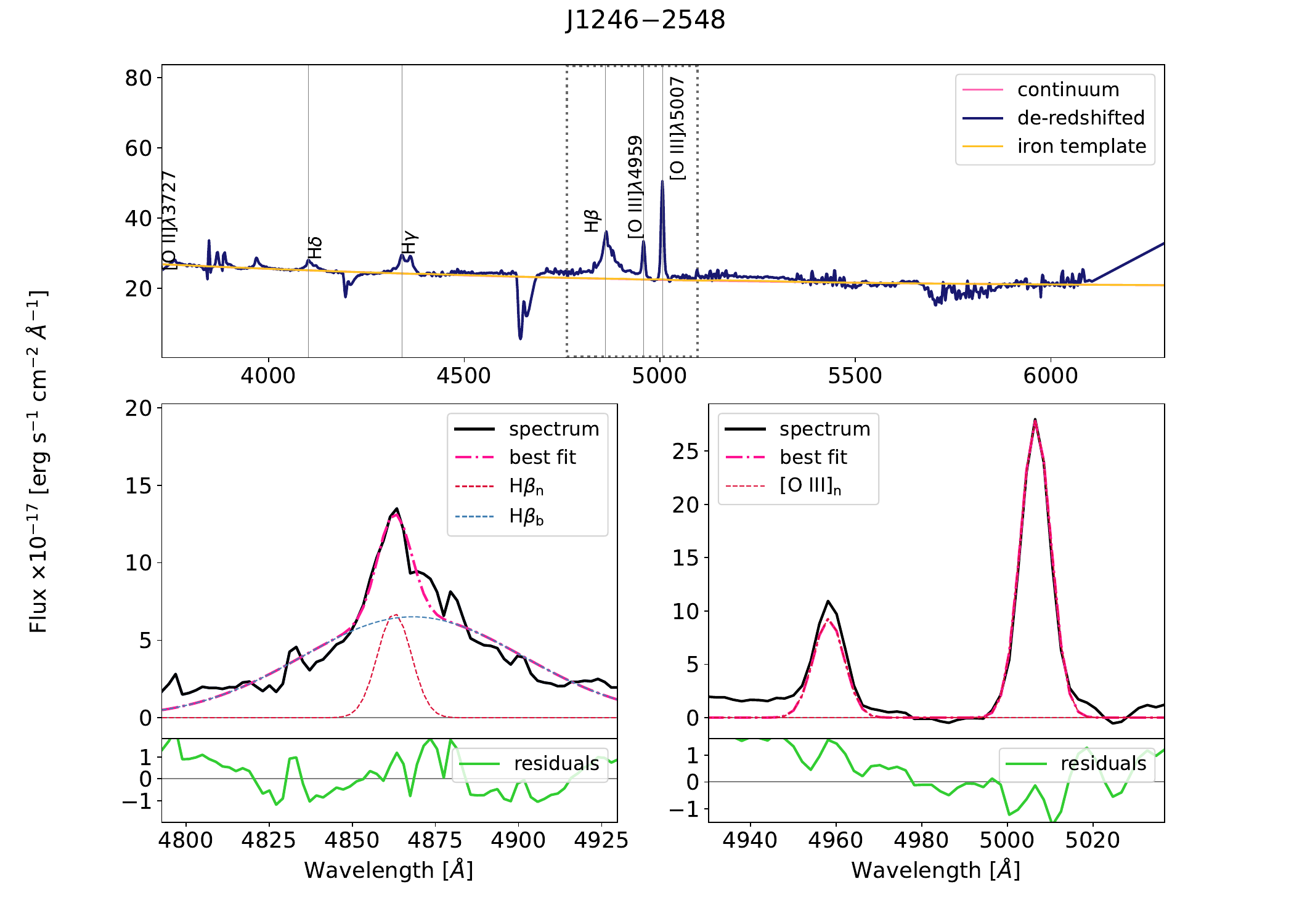} 
      \includegraphics[width=0.44\textwidth]{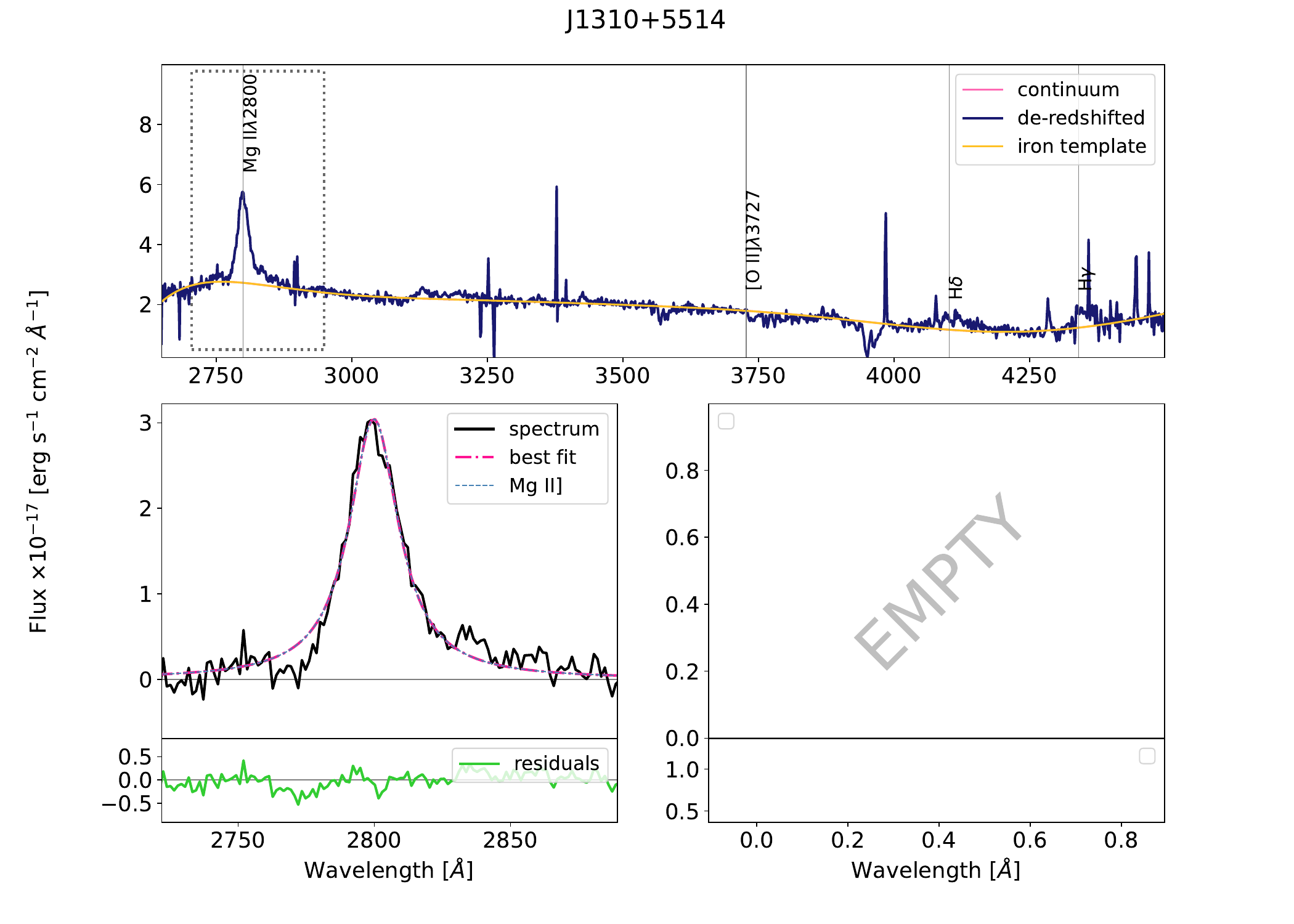}
     \includegraphics[width=0.44\textwidth]{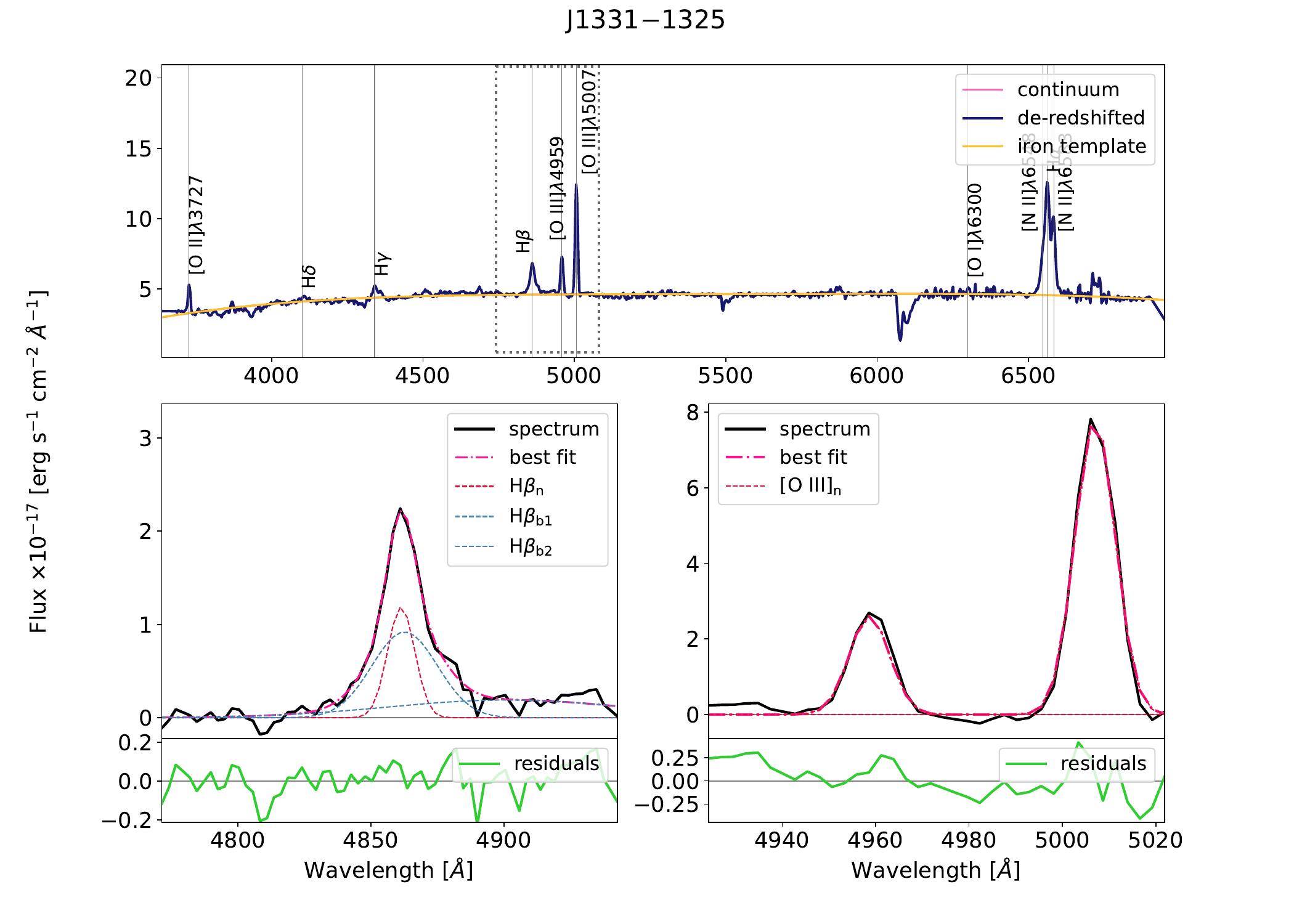}
          \end{figure*}
\begin{figure*}
    \centering
    \includegraphics[width=0.44\textwidth]{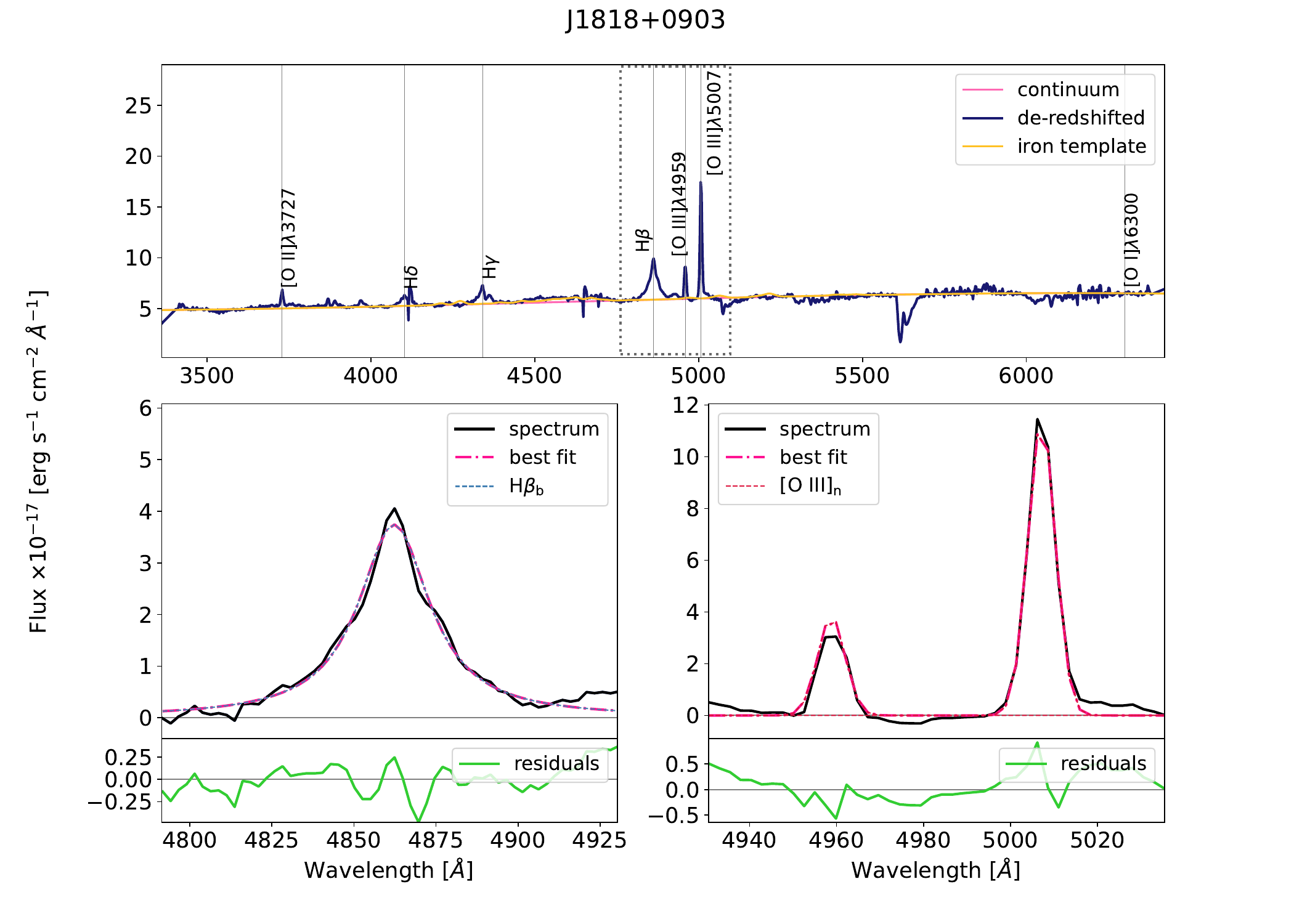}
    \includegraphics[width=0.44\textwidth]{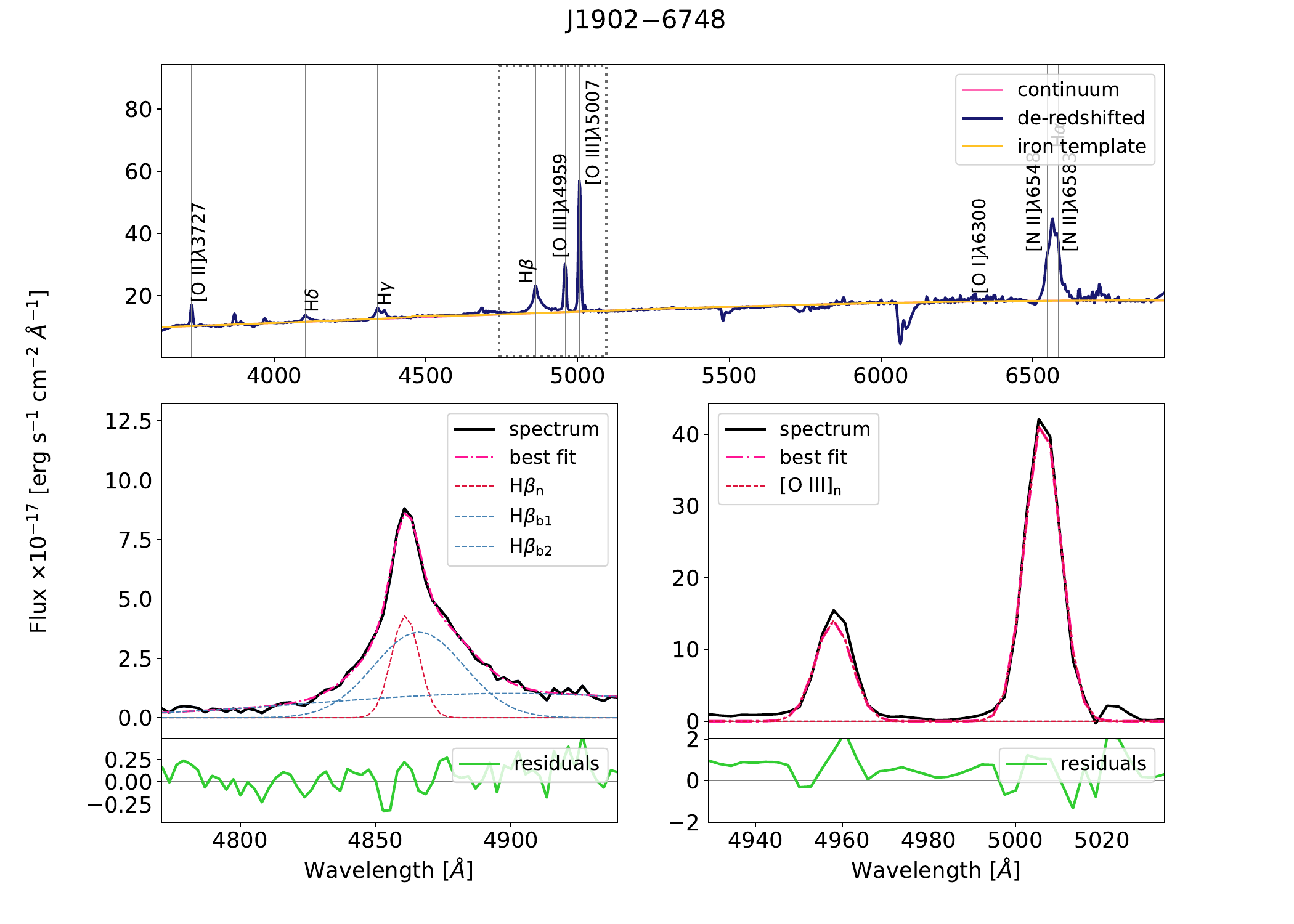} 
    \includegraphics[width=0.44\textwidth]{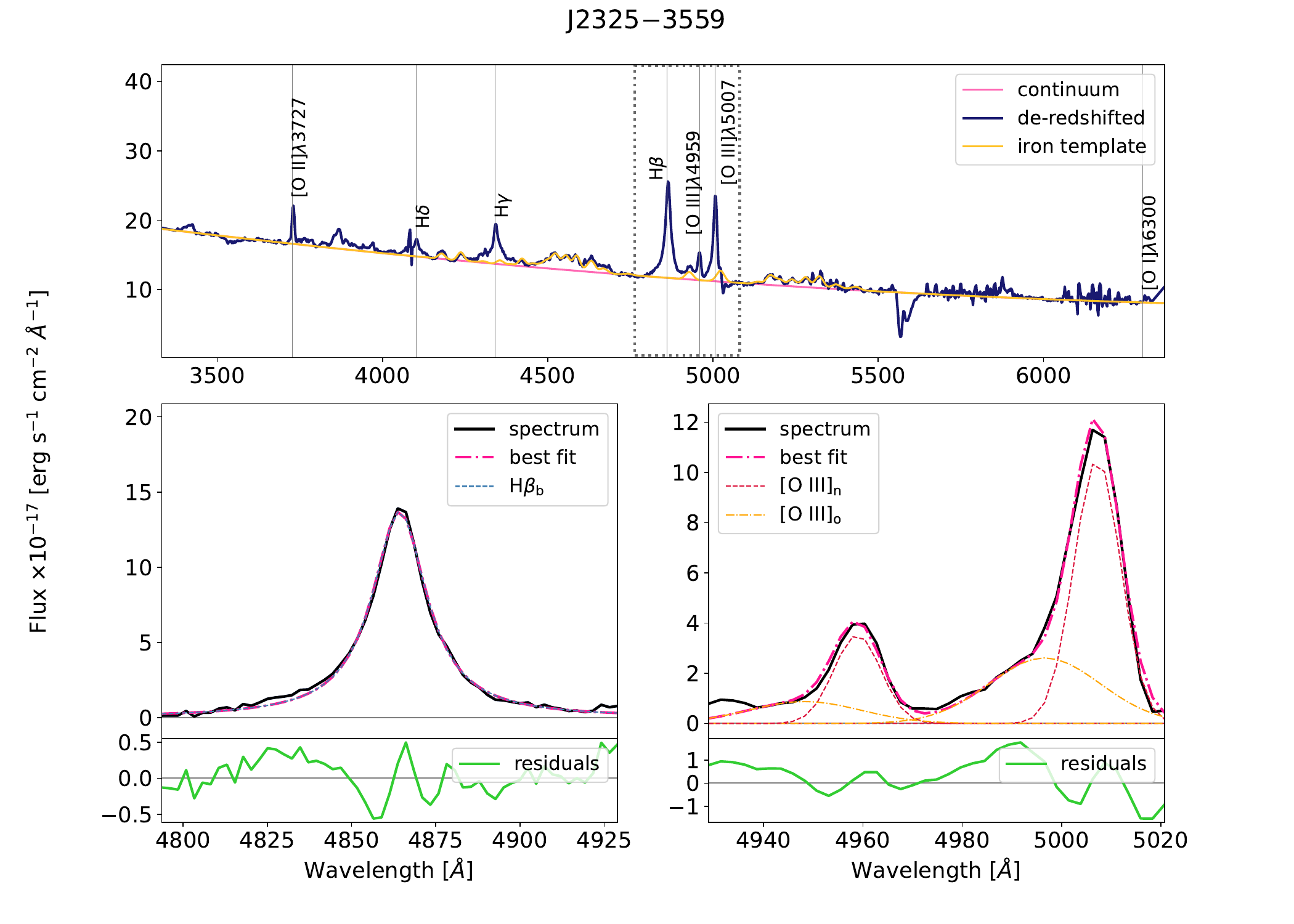} 
    \includegraphics[width=0.44\textwidth]{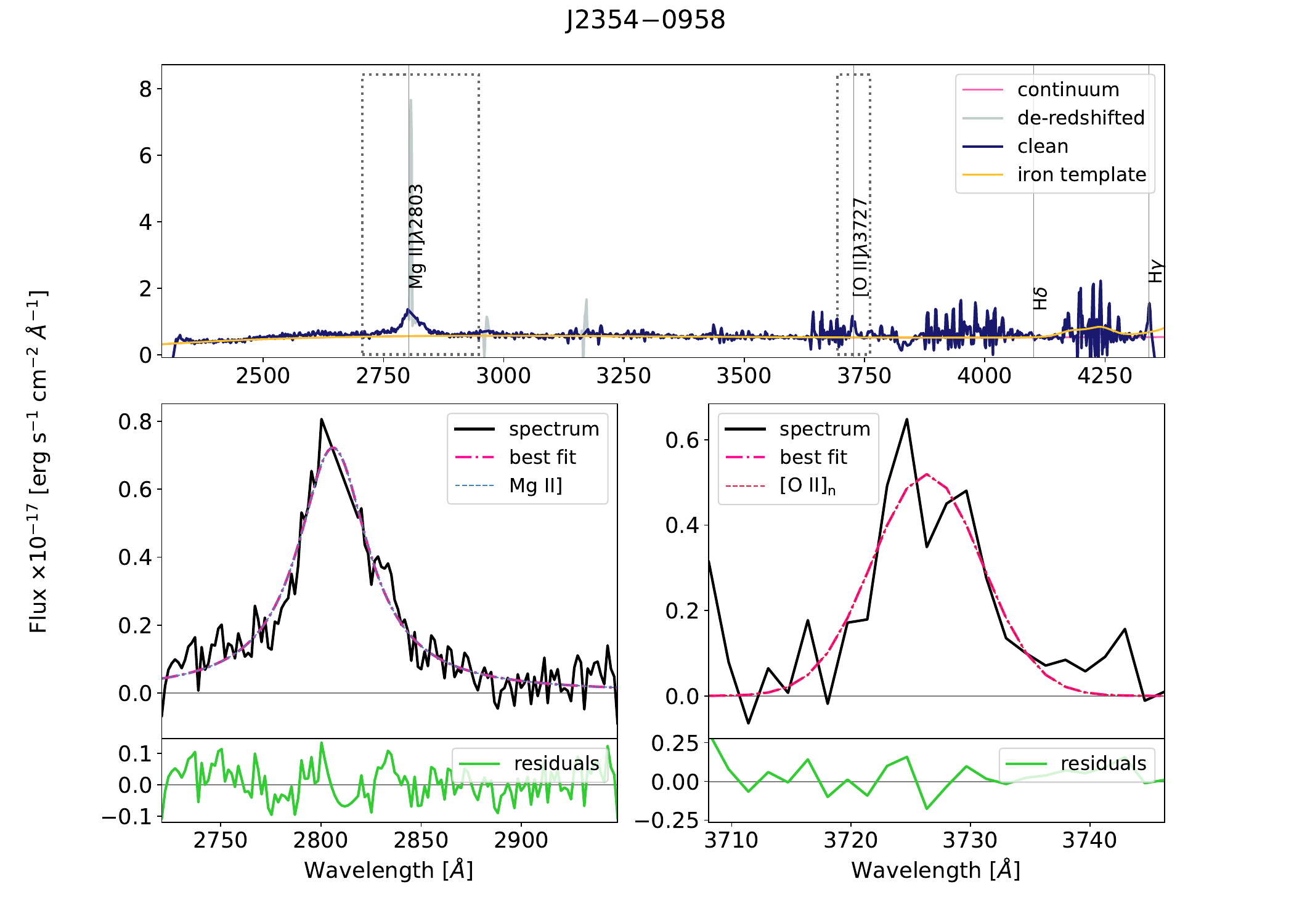} 
\end{figure*}

\FloatBarrier

\section{List of the optically confirmed $\gamma-$ray detected NLS1s}\label{appendix_gamma_nls1}
\begin{table}[h!]
\centering
\caption{List of $\gamma-$NLS1s with the corresponding coordinates and the most recent reference for the classification.}
	\begin{tabularx}{\textwidth}{XlXXXXl}
            \hline
		 {\bf Name} &  {\bf Alias} &  {\bf RA (J2000)} &  {\bf DEC (J2000)} & {\bf z} & {\bf Classification} \\
		 {\bf } &  {\bf } &  {\bf [h:m:s]} &  {\bf [$^{\circ}$:':'']} & & {\bf reference} \\
            \hline
            \hline
            	J0001$+$2113	&	TXS 2358$+$209			& 00:01:32.4 	&   $+$21:13:36.3	&	0.438	&	a \\ 
		J0010$+$2043	&	TXS 0007$+$205			& 00:10:28.7	&   $+$20:47:49.8	&	0.597	&	a \\ 
		J0014$-$0500	&	WISE J001420.42$-$045928.7	& 00:14:20.4	&   $-$04:59:28.8	&	0.790	&	a \\
		J0038$-$0204$^{\dagger}$&	3C 17			& 00:38:20.5	&   $-$02:07:40.5	&	0.220	&	a \\
		J0102$+$4214	& 	GB6 J0102$+$4214    		& 01:02:27.1	&   $+$42:14:18.9	&	0.877	&	b \\ 
		J0105$+$1912$^{\dagger}$&	TXS 0103$+$189	& 01:05:55.2	&   $+$19:12:28.0	&	0.879	&	a \\
		J0224$+$0700 	&    	PKS 0221$+$067     			& 02:24:28.4  	&   $+$06:59:23.3	&	0.512	&	b \\         
		J0324$+$3412	&	1H 0323$+$342			& 03:24:41.2 	&   $+$34:10:45.9	&	0.063	&	b \\
		J0422$-$0644	&	PMN J0422$-$0643    		& 04:22:10.8   	&   $-$06:43:45.3	&     	0.242	&	c \\
		J0850$+$5108	&	SBS 0846$+$513			& 08:49:58.0	&   $+$51:08:29.0	&	0.584	&	a \\ 
		J0932$+$5306 	&    	S4 0929$+$53       			& 09:32:41.2     &   $+$53:06:33.8	&	0.597	&	c \\         
		J0933$-$0013$^{\dagger}$&	PMN J0933$-$0012	& 09:33:23.0	&   $-$00:10:51.6	&	0.796	&	a \\
		J0948$+$0022	&	PMN J0948$+$0022			& 09:48:57.3 	&   $+$00:22:25.6	&	0.585	&	d \\ 
		J0949$+$1749$^{\dagger}$& TXS 0946$+$181	& 09:49:39.8 	&   $+$17:52:49.4	&	0.692	&	a \\
		J0958$+$3222	&	3C 232					& 09:58:20.9 	&   $+$32:24:02.2	&	0.530	&	a \\
		J0959$+$4600	&	SDSS J095909.51$+$460014.3 & 09:59:09.5 	&   $+$46:00:14.3	&	0.399	&	e \\  
		J1127$+$3618$^{\dagger}$&	MG2 J112758$+$3620& 11:27:58.9 	&   $+$36:20:28.4	&	0.884	&	a \\ 
		J1202$-$0528  	&    	PKS 1200$-$051    			& 12:02:34.2     &   $-$05:28:02.5	&	0.381	&	c \\       
		J1208$+$5441	&	TXS 1206$+$549			& 12:08:54.3 	&   $+$54:41:58.2	&	1.34		&	b \\
		J1214$-$1926	&	PKS B1211$-$190			& 12:14:03.4 	&   $-$19:21:42.8	&	0.149	&	b \\
		J1246$-$2548	&	PKS 1244$-$255			& 12:46:46.8	&   $-$25:47:49.3	&	0.637	&	c \\
		J1305$+$5118	&	IERS B1303$+$515			& 13:05:22.8 	&   $+$51:16:40.3	&	0.787	&	a \\
		J1310$+$5514 	&    	TXS 1308$+$554     			& 13:11:03.2     &   $+$55:13:54.3	&	0.926	&	b \\        
		J1443$+$4728	&	B3 1441$+$476			& 14:43:18.6 	&   $+$47:25:56.7	&	0.703	&	a \\
		J1505$+$0326	&	PKS 1502$+$036			& 15:05:06.5 	&   $+$03:26:30.8	&	0.407	&	a \\
		J1520$+$4209$^{\dagger}$&	B3 1518$+$423	& 15:20:39.7	&   $+$42:11:11.5	&	0.485	&	a \\ 
		J1639$+$4129&	MG4 J163918$+$4127		& 16:39:15.8	&   $+$41:28:33.7	&	0.690	&	a \\
		J1644$+$2620	&	MG2 J164443$+$2618		& 16:44:42.5 	&   $+$26:19:13.3	&	0.144	&	a \\
		J1700$+$6830	&	TXS 1700$+$685			& 17:00:09.3 	&   $+$68:30:07.0	&	0.301	&	b \\
		J1818$+$0903	&   	NVSS J181840$+$090346 	& 18:18:40.1    	&   $+$09:03:46.2	&	0.354	&	c \\         
		J1848$+$3217	&	B2 1846$+$32A			& 18:48:22.1 	&   $+$32:19:02.6	&	0.798	&	b \\
		J2118$+$0019$^{\dagger}$&	PMN J2118$+$0013	& 21:18:17.4 	&   $+$00:13:16.8	&	0.463	&	a \\
		J2118$-$0723$^{\dagger}$&	TXS 2116$-$077	& 21:18:53.0	&   $-$07:32:27.6	&	0.260	&	a \\
		J2325$-$3559  	&    	CTS 0490      				& 23:25:28.6 	&   $-$35:57:54.2	&	0.367	&	c \\        
            \hline
	\end{tabularx}
	\tablefoot{References: (a) \citet{2024MNRAS.527.7055P}, (b) \citet{2022Univ....8..587F}, (c) this work, (d) \citet{2025A&A...698A.320D}, and (e) \citet{2023A&A...676A...9L}. Objects marked with $^{\dagger}$ were originally classified as $\gamma$-NLS1s by \citet{2024MNRAS.527.7055P}, but after visual inspection of the SDSS spectra we suggest their reclassification as ambiguous, due to spectral features or poor data quality. Additional support for this ambiguity comes from the case of J2118$-$0723, which was reclassified from NLS1 to IS by \citet{2020A&A...636L..12J} after a detailed analysis.}
	\label{tab_gamma_nls1}
\end{table}

\end{appendix}

\end{document}